\documentclass[12pt]{iopart}
\usepackage{cite}
\usepackage{graphicx}
\usepackage{amssymb}
\DeclareGraphicsExtensions{.eps}
\usepackage[cmex10]{amsmath}
\usepackage{array}
\usepackage{subfig}
\graphicspath{{./paper_figures/}}
\usepackage[author={Asheq}]{pdfcomment}
\usepackage[linesnumbered,resetcount,lined,boxed,commentsnumbered,ruled,vlined,longend]{algorithm2e}
\SetAlFnt{\footnotesize\sf}

\def\w{w}
\def\r{\boldsymbol{r}}
\def\s{\hat{\boldsymbol{s}}}

\def\n{\hat{\boldsymbol{n}}}
\def\rs{\boldsymbol{r}_{\mathrm{s}}}
\def\E{{\mathcal{E}}}
\def\K{\mathcal{K}}
\def\X{\mathcal{X}}

\def\bA{\mathbf{A}}
\def\bAhat{\mathbf{\hat{A}}}
\def\A{\mathcal{A}}
\def\Ahat{\hat{\mathcal{A}}}
\def\cm{c_\mathrm{m}}
\def\mm{\mathrm{mm}}

\def\th{\mathrm{th}}
\def\eff{\mathrm{eff}}
\def \emis{\boldsymbol{\lambda}}
\def \att {\boldsymbol{\mu}}
\def \attn {\boldsymbol{\mu}}

\def \sc {\mathrm{sc}}

\def \L {\mathcal{L}}

\def \dr{\mathrm{d}}
\def \Q{\mathcal{Q}}

\def \P {\mathbb{P}}
\def \path{\mathbb{P}}
\def \pathp{\mathbb{P'}}

\def \E{E}

\def \magni{\mathrm{mag}}
\def \subpath{\mathbb{S}}
\def \pr{\mathrm{pr}}
\def \Pr{\mathrm{Pr}}

\def \cmeter{\mathrm{cm}}

\def \Kdet{\boldsymbol{K}_{\mathrm{det}}}
\def \keV{\mathrm{keV}}

\usepackage{xcolor}
\usepackage{soul}

\let\oldnl\nl% Store \nl in \oldnl
\newcommand{\nonl}{\renewcommand{\nl}{\let\nl\oldnl}}
\newcommand{\nextnr}{\stepcounter{AlgoLine}\ShowLn}

\begin{document}

\title[Fisher info. analy. for joint act.- attn. recon. using LM SPECT] {Fisher information analysis of list-mode SPECT emission data for joint estimation of activity and attenuation distribution}
%\author{{Abhinav~K.~Jha$^{1,2}$, Yansong Zhu$^{3,4}$, Matthew~A.~Kupinski$^5$ and Eric~C.~Frey$^3$}
\author{{Md Ashequr Rahman$^{1,2}$, Yansong Zhu$^{3,4,5}$, Eric~Clarkson$^6$, Matthew~A.~Kupinski$^6$, Eric~C.~Frey$^3$ and Abhinav~K.~Jha$^{1,2}$}
\address{$^1$Department of Biomedical Engineering, Washington University in St. Louis, St. Louis, MO, USA}
\address{$^2$Mallinckrodt Institute of Radiology, Washington University in St. Louis, St. Louis, MO, USA}
\address{$^3$Department of Radiology and Radiological Sciences, Johns Hopkins University, Baltimore, MD, USA}
\address{$^4$Department of Electrical and Computer Engineering, Johns Hopkins University, Baltimore, MD, USA}
\address{$^5$Department of Physics $\&$ Astronomy, University of British Columbia, Canada}
\address{$^6$College of Optical Sciences, University of Arizona, Tucson AZ, USA.}
}
\maketitle
%\IEEEpeerreviewmaketitle

\begin{abstract}
The potential to perform attenuation and scatter compensation (ASC) in single-photon emission computed tomography (SPECT) imaging  without a separate transmission scan is highly significant.
In this context, attenuation in SPECT is primarily due to Compton scattering, where the probability of Compton scatter is proportional to the attenuation coefficient of the tissue and the energy of the scattered photon and the scattering angle are related.
Based on this premise, we investigated whether the SPECT scattered-photon data acquired in list-mode (LM) format and including the energy information can be used to estimate the attenuation map. 
For this purpose, we propose a Fisher-information-based method that yields the Cramer-Rao bound (CRB) for the task of jointly estimating the activity and attenuation distribution using only the SPECT emission data.
In the process, a path-based formalism to process the LM SPECT emission data, including the scattered-photon data, is proposed. The Fisher information method was implemented on NVIDIA graphics processing units (GPU) for acceleration.
The method was applied to analyze the information content of SPECT LM emission data, which contains up to first-order scattered events, in a simulated SPECT system with parameters modeling a clinical system using realistic computational studies with 2-D digital synthetic and anthropomorphic phantoms. The method was also applied to LM data containing up to second-order scatter for a synthetic phantom.
Experiments with anthropomorphic phantoms simulated myocardial perfusion and dopamine transporter (DaT)-Scan SPECT studies. 
The results show that the CRB obtained for the attenuation
and activity coefficients was typically much lower than the true value of these coefficients.
An increase in the number of detected photons yielded lower CRB for both the attenuation and activity coefficients.  
Further, we observed that systems with better energy resolution yielded a lower CRB for the attenuation coefficient.
%We also observed that for the first-order scatter case, the emission data in LM format yielded a lower CRB in comparison to binned data.
Overall, the results provide evidence that LM SPECT emission data, including the scattered photons, contains information to jointly estimate the activity and attenuation coefficients.
\end{abstract}


\section{Introduction}
In single-photon emission computed tomography (SPECT) imaging, a radiotracer that emits gamma-ray photons is injected into the patient.
From the detected gamma-ray photons, the radiotracer distribution within the patient is reconstructed. 
However, a fraction of photons scatter as they propagate through the tissue leading to scatter and attenuation artifacts.
Thus, compensation of scatter and the resultant attenuation, referred to as attenuation and scatter compensation (ASC), is required for reliable reconstruction. 
ASC is a prerequisite for absolute quantification of the tracer uptake and has been observed to benefit visual-interpretation tasks  \cite{Hutton_2011, Garcia:07, Bailey:13, Masood}.
To perform ASC, an attenuation map of the patient is required. 
Conventional ASC methods obtain this map using a transmission scan, typically a CT scan of the patient.  
However, these CT-based ASC methods suffer from many issues such as higher costs, increased radiation dose, and  possibility of misregistration between the SPECT and CT scans, which could lead to inaccurate diagnosis \cite{Hutton_2011, Connor, Germano:07, Stone:98, He:10}.
Current commercial scanners that perform ASC are often dual-modality SPECT/CT systems, which are substantially more expensive than conventional SPECT systems and often require larger imaging rooms, additional shielding, and more complicated acquisition protocols.
Further, currently, a majority (around 80\%) of the SPECT market share is occupied by stand-alone SPECT systems \cite{Technavio:17}. %TODO
Additionally, several emerging solid-state-detector-based SPECT systems, which have demonstrated capability to provide images at low dose, do not have CT imaging capability \cite{Liu:14, Caobelli:16, Palyo:16}. %TODO
%Due to all these reasons, the potential to estimate the attenuation distribution without using a transmission scan and using only the SPECT emission data would be highly significant \cite{Hutton_2011}.
%The importance of such a method is magnified given that stand-alone SPECT systems comprise a majority (around 80 \%) of the SPECT market share. 
Due to all these reasons, a method that estimates the attenuation map using only the SPECT emission data is poised to have a strong impact on the SPECT imaging landscape \cite{Hutton_2011}. 
Given this high significance, in this manuscript, we address the inverse problem of jointly estimating the activity and attenuation distribution using only the SPECT emission data. 

Existing techniques for estimation of the attenuation map from SPECT emission data can be divided in two classes.
The first class of methods uses the scattered data to reconstruct the attenuation images using simple methods such as filtered back-projection (FBP) \cite{Pan_1996, Pan_1997, Nunez, Michel, Zaidi_2003}.
These methods use the fact that Compton scattering is the dominant photon-interaction mechanism in soft tissue, and the probability of Compton scatter is directly proportional to the attenuation coefficient.
Thus, the reconstructed images could show contrast between tissues with different attenuation coefficients. 
The different regions can be segmented in these images and pre-defined attenuation coefficients can be assigned to these regions. 
These methods work reasonably well when the activity is widely distributed, but have limitations when the activity is concentrated within a small-sized region \cite{Hutton_2011}.
%These methods are also not theoretically rigorous. 
Further, assuming known attenuation coefficients can be inaccurate in organs such as lungs where the density varies depending on several factors including disease state. 
The second class of methods estimate the attenuation coefficients directly from the emission data. 
These algorithms either perform iterative inversion of the forward mathematical model \cite{Censor, Nuyts,Krol,Gourion}, or exploit the consistency conditions based on the forward model\cite{Natterer, Welch,Yan,Crepaldi}. 
However, most of these methods are slow and neglect scattered photons. The techniques have met with limited success \cite{Hutton_2011}.

A more recent study explored the potential of inverting the models used for scatter to estimate the attenuation distribution \cite{Cade_2013}. 
The study was limited in terms of considering only two energy windows, binned data, and two-dimensional (2D) phantoms. 
However, even with these limitations, it was observed that different regions of attenuation were distinguished for physical phantoms. 
The reconstruction results were not very accurate, and the computation time was high, but as the authors commented, it was a promising first step.
Of most importance, this study showed that inverting models used for scatter can help estimate the attenuation distribution. Similar inversion-based methods \cite{cueva2018algebraic,courdurier2015simultaneous,Bousse:16} have shown potential for using scattered data for simulataneous reconstruction of activity and attenuation coefficients.

\begin{figure}
\centering
\includegraphics[width = 4 in]{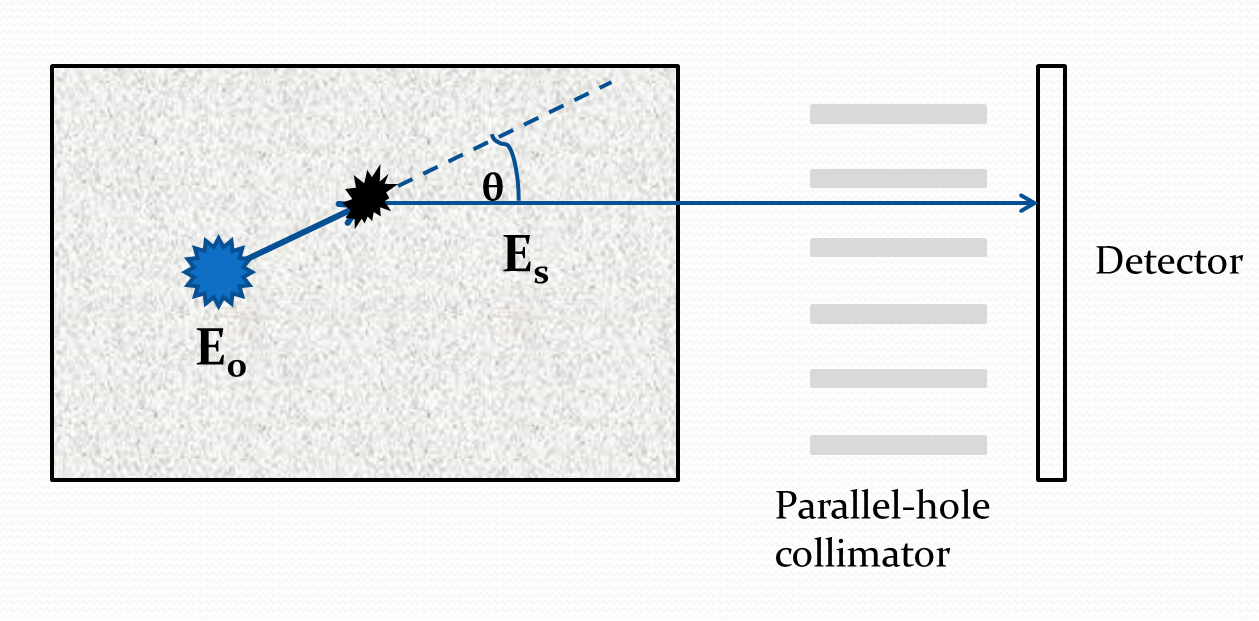}
\caption{A schematic illustrating the intuition behind how the energy of the scattered photon can help determine the scattering location. Due to the relation between the angle of scatter and energy of scattered photon, in a hypothetical scenario with known point source and ideal collimator and detector, the location of scatter can be determined.}
\label{fig:intution_energy_attenuation_information}
\end{figure}

In SPECT imaging, for each detected photon, several attributes such as the position of interaction, energy deposited, and time of interaction can be estimated. 
The energy deposited by the scattered photon and the angular orientation of the detector can yield information about the location of scatter.
To illustrate this intuitively in an ideal scenario, consider a 2D SPECT imaging system that has perfect energy and position resolution and only allows photons perpendicular to the detector face (Fig.~\ref{fig:intution_energy_attenuation_information}). 
In this case, for any detected photon, the energy and the direction of the detected photon will be precisely known. 
We know that there is a direct relationship between the angle of scatter and the energy deposited by the scattered photon. 
Thus, for an ideal system, the scatter angle can be precisely computed using the energy attribute of the scattered photon. 
In that case, if we assume that the emission source is a point source whose location is known, then we could precisely determine the location at which the photon scattered, as illustrated in Fig.~\ref{fig:intution_energy_attenuation_information}. This information regarding scattering location can then help to estimate the attenuation coefficient from the detected LM events. 

The above-described methods to estimate the attenuation map from the SPECT emission data do not explicitly use this energy attribute. 
Further, for each detected photon, all the attributes can be stored in a list-mode (LM) format \cite{Barrett_1997}.
In the above-described methods, the attribute space is instead discretized into bins, and a given photon is allotted to a bin based on its attribute value.
%For example, a photon of energy 124.3 keV could be assigned to the bin corresponding to scatter energy window between, let us say, 80 and 126 keV.
As expected, the binning operation leads to information loss.
%Mathematically, the binning operation maps the continuous object space into a finite-dimensional data space, which implies that an infinite-dimensional null space is gauranteed due to dimensionality considerations \cite{Jha:15:pmb}.
%However, when we process the data in list-mode format without binning, then the 
The effect of binning-related information loss when the photon attributes of position and time of interaction are binned has been shown on the null functions \cite{Jha:15:pmb} and on quantification \cite{Jha:15:spie} in a SPECT system.  
%It has been observed that systems that bin the photon attributes measure a smaller set of object features in comparison to systems that do not bin the attributes and instead process data in LM format\cite{Jha:15:pmb}.
%Further, the systems have process the attributes in LM format also yield improved estimation of activity uptake \cite{Jha:15:spie}.

Our manuscript is motivated by previous studies on inverting the models used for scatter \cite{Cade_2013,cueva2018algebraic,courdurier2015simultaneous,Bousse:16}, on the information loss that is avoided by processing data in LM format \cite{Jha:15:pmb, Jha:15:spie}, and the potential that the energy attribute contains information about the scattering coefficient.
We investigate whether the SPECT emission data, including the scattered photons, processed in LM format and including the energy attribute, can jointly estimate the activity and attenuation distribution by inverting the models used for scatter. 
For this purpose, we develop a novel Fisher information-based method that quantifies the information content in LM SPECT emission data for  jointly estimating the activity and attenuation distribution. 
The method requires processing SPECT emission data, including the scattered photons, in LM format. 
We propose a new path-based formalism for this purpose. 
Application of the proposed Fisher-information-based method to computational studies yields several novel insights about the information content in scattered photons in SPECT. 
Preliminary versions of the theoretical treatment described in this paper have been presented previously \cite{Jha:13:spie, Jha:13:diss}.

\section{Theory}
\subsection{Problem formulation and a path-based formalism to process scattered-photon data}
\label{sec:path_formalism}
Consider a SPECT imaging system consisting of scintillation cameras that is detecting gamma-ray photons emitted by an object. This object consists of an activity distribution and an attenuation distribution, which parameterize the emission and attenuation of photons, respectively. Consider that the object being imaged is represented in a voxel basis consisting of $N$ voxels, so that the activity and attenuation distributions are a set of $N$-dimensional vectors, denoted by $\emis$ and $\att$, respectively.
Denote the $N$ elements of the activity and attenuation distribution vector, by $\{\lambda_1, \ldots , \lambda_N \}$ and $\{\mu_1, \ldots, \mu_N \}$, respectively, where $\lambda_i$ is the mean number of photons emitted from the $i^{\th}$ voxel per unit time, and $\mu_i$ is the attenuation coefficient of the $i^{\th}$ voxel. When a transmission scan is used to measure the attenuation distribution $\att$, then the inverse problem is to estimate $\emis$. However, in this paper, we are considering that a transmission scan is unavailable, so the inverse problem is jointly estimating $\emis$ and $\att$ from only the SPECT emission data. 
 
For each gamma-ray photon emitted from the object that interacts with the scintillator, the position of interaction of the gamma-ray photon with the crystal and the energy deposited at the interaction site are estimated and recorded.
Denote the true and estimated attributes of the $j^{\th}$ detected event by the attribute vectors $\bA_j$ and $\bAhat_j$, respectively.
The attribute vector is a $q$-dimensional vector where $q$ is the number of attributes in each LM event. We assume that the number of detected counts $J$  is fixed, i.e.~we have a preset-count system. Note that our analysis is general and can be easily extended to a preset-time system. 
Denote the full LM dataset of estimated attributes as $\Ahat = \{ \bAhat_j, j = 1,2,...,J \}$.

We can represent the LM data as an impulse-valued random process on attribute space \cite{barrett2013foundations}, given by the generalized function:
\begin{align}
g({\bA})=\sum_{j=1}^{J}\delta({\bA}-\bAhat_j) .
\end{align}
As mentioned above, this data can be binned, and for completeness, we present the mathematical representation of the binned data. A bin can be defined as a hyperrectangle in the attribute space, denoted by the function $b_m(\bA)$ for $m=1,...M$. Here we assume that the binning
functions represent non-overlapping hyperrectangle and encompass the full range of
attribute space. In that case, the measurement in the $m^{\th}$ bin, denoted by $g_m$ is given by \cite{clarkson2019quantifying}
\begin{align}
g_m = \int g(\bA)b_m(\bA)d^q\bA = \sum_{j=1}^{J}b_m(\bAhat_j),
\end{align}
where the integral is over the attribute space. We observe that the binned data is the result of an additional binning function applied to the LM data, and thus intuitively is expected to lead to  an information loss, as also mentioned above. This has also been observed in previous studies \cite{Jha:15:pmb,Jha:15:spie, clarkson2019quantifying}. Thus, we focus our analysis on jointly estimating the activity and attenuation distribution directly from the LM data. 

More specifically, we intend to quantitatively determine if the LM data contains information to jointly estimate $\emis$ and $\attn$. We use the widely used metric of Fisher information to quantify this information content \cite{barrett2013foundations, van2004detection}.  Deriving the Fisher information requires computing an expression for the likelihood. This is given by\cite{Barrett_1997}
\begin{equation}
\pr(\hat{\A}, T | \emis, \att ) = \pr(T|\emis, \att) \prod_{j=1}^{J} \pr(\bAhat_j | \emis, \att),
\label{likelihood}
\end{equation}
where $T$ is the acquisition time required for detecting $J$ LM events.
%\pdfcomment{modifications done: please let me know if I should provide any other details here.}
Taking the logarithm on both sides of the Eq.~\eqref{likelihood} yields the log-likelihood of the observed LM data, denoted by $\L( \emis, \att | \hat{\A}, T)$ and given by
\begin{equation}
\L( \emis, \att | \hat{\A}, T) = \sum_{j=1}^J \log \pr(\bAhat_j |  \emis, \att  ) + \log \pr(T | \emis, \att).
\label{loglikelihood1}
\end{equation}
To quantify the Fisher information in the LM data for jointly estimating $\emis$ and $\attn$, the log-likelihood must be differentiated with respect to the elements of $\emis$ and $\attn$. 
This requires obtaining an analytic expression for $\pr(\bAhat_j | \emis, \att  )$ and $\pr(T | \emis, \att )$.
For a fixed number of detected counts $J$, the acquisition time $T$ follows an Erlang distribution with shape $J$ and rate $\beta$, where $\beta$ is the mean rate of photons deected by the detector \cite{Parra_1998}, so that
\begin{equation}
\pr(T| \emis, \att) = \frac {\beta^{J} T^{J-1}\exp(-\beta T) }{(J-1)!}.
\label{prob_time}
\end{equation}
Obtaining a similar direct analytic expression for $\pr(\bAhat_j | \emis, \att )$ is complicated.
To address this issue, note that any LM event is the result of a photon emitted from a voxel, travelling in a certain direction, and then, in some cases, scatterring in certain voxels and finally being detected by the detector.
In other words, any LM event is a result of a photon traveling within a discrete unit of space, which we refer to as a path. 
%This is illustrated for three different LM events in Fig.~\ref{scat_path}.
%A path $\P$ taken by a gamma-ray photon can be defined by a set of voxels $\{k_0, k_1, \ldots k_n \}$, where $k_0$ is the voxel from which the photon is emitted, and $k_1, k_2, \ldots k_n$ denote the voxels in which the photon scatters.
%More rigorously, a path is a discrete unit with a finite angular range that connects different voxels through which a photon travels. 
The concept of a path will be mathematically defined in the next section, but for now, suffice to say that a path is a discrete variable, denoted by $\path$.   
The expressions for the PDF of the LM event given the path, $\pr(\bAhat_j | \P)$, and the probability mass function of the path, $\Pr(\P | \emis,\att)$, can be derived, as described later. 
We thus decompose $\pr(\bAhat_j |  \emis, \att  )$ as a mixture model over all possible paths.
For this purpose, we use the following identity:
\begin{equation}
\pr(x|y=Y) = \sum_z \pr (x | y=Y, z=Z) \Pr (z=Z| y=Y), 
\end{equation}
where $x$ denotes a continuous random variable, and $y$ and $z$ denote discrete random variables. 
In the considered scenario, $x$, $y$ and $z$ correspond to the LM event attributes, emission and attenuation vectors, and the path, respectively.
Applying this identity yields the following mixture model for $\pr(\bAhat_j | \emis, \att )$:  
\begin{equation}
\pr(\bAhat_j |  \emis, \att) = \sum_{\P} \pr(\bAhat_j | \P, \emis, \att) \Pr(\P | \emis, \att).
\label{mixture_model}
\end{equation}
The components of the mixture model are the probabilities that a LM event occurs given the photon traces a path, and the weight of each component is the probability of the considered path.
Because the LM event has already occurred, the probability of the event given the path is independent of the activity and attenuation distribution, i.e.~$\pr(\bAhat_j | \P, \emis, \att) = \pr(\bAhat_j | \P)$, provided the probability of the path accounts for the emission and attenuation processes, as will be the case in our treatment. 
Using Eq.~\eqref{loglikelihood1}, we can rewrite the log-likelihood of the data given the activity and attenuation distribution in terms of this mixture-model decomposition as
\begin{equation}
\L( \emis, \att | \hat{\A}, T) = \sum_{j=1}^J \log \sum_{\P} \pr(\bAhat_j | \P) \Pr(\P | \emis, \att) +  \log \pr(T | \emis, \att).
\label{loglikelihood2}
\end{equation}
To derive the elements of the FIM, analytic expressions for $\Pr(\P | \emis, \att)$ and $\pr(\bAhat_j | \P)$ must be derived. These are the topics of the next two sub-sections. 

%\begin{figure}
%\centering
%\includegraphics[width = 2 in]{scat_path.png}
%\caption{A schematic showing the various paths that a emitted gamma-ray photon can take leading to different sets of LM events.}
%\label{scat_path}
%\end{figure}

\subsection{Computing radiation transfer through a path}
\label{sec:rad_transfer_path}

In this section, we derive the expression for $\Pr(\P | \emis, \att)$. 
We first mathematically define a path. 
A path is a discrete unit of space that connects different voxels through which photon radiation propagates.
A path can be described in terms of a set of sub-paths, where a sub-path describes the unit of space through which radiation propagates between two voxels.
To describe the radiation transfer through a sub-path, we use an approach similar to the discrete-ordinates method for solving the equation of radiative transport \cite{Bouaoun:05}. 
Each sub-path is defined in terms of a start location and a finite angular range.
First, assume that the directional coordinates are discretized by dividing the angular space of $4 \pi$ radians into a finite number of  solid-angle sub-domains, referred to as ordinates, each of size $\Delta \Omega$ . 
Denote the 3-D unitary direction vector by $\s$ and discretized angular direction by $\s_k$ associated with the $k^{\th}$ instance of the ordinate. 
Then the indicator function of the $k^{\th}$ angular ordinate $\psi_k(\s)$ is defined as below:
\begin{equation}
\psi_k(\s) = \begin{cases}
1, \quad\text{if $k^{\th}$ ordinate contains the direction vector $\s$.} \\
0, \quad\text{otherwise.}
\end{cases}
\end{equation}

\begin{figure}
\centering
\includegraphics[width = 4 in]{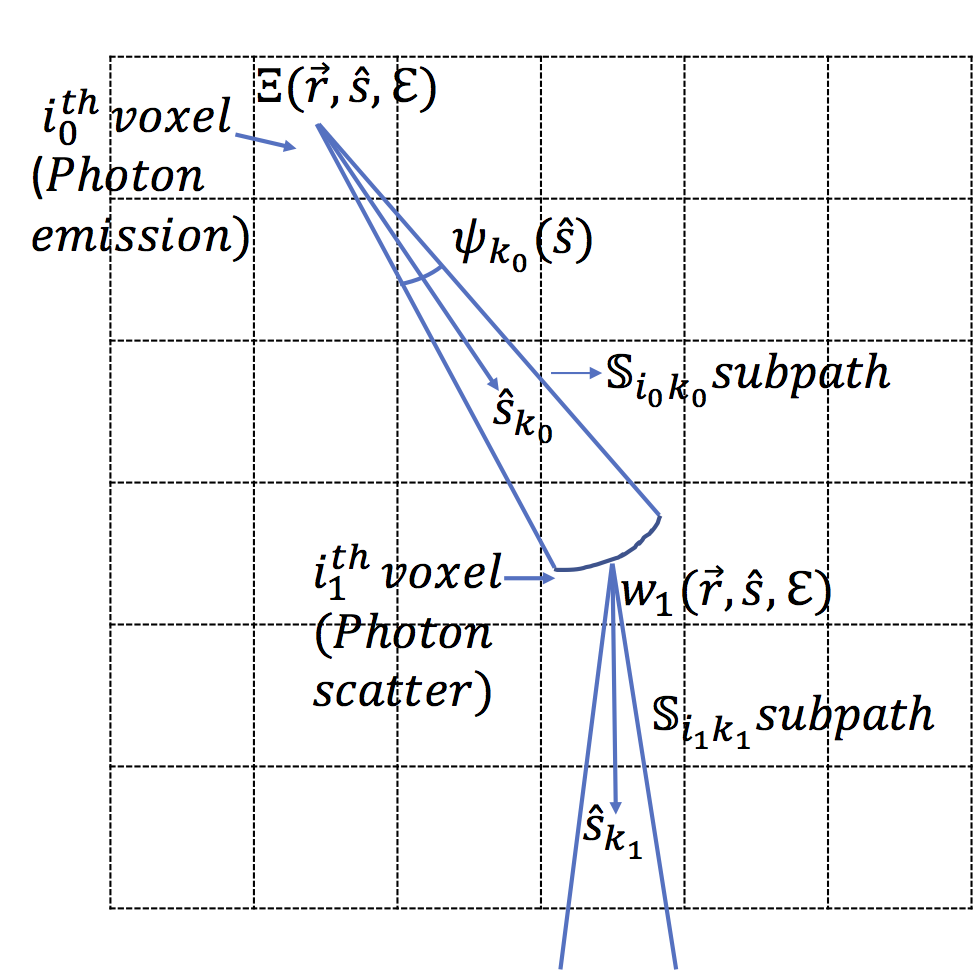}
\caption{A schematic summarizing the various notations used to describe radiation transfer through a path for a 2-D setup.}
\label{fig:notation_path}
\end{figure}

A sub-path $\subpath_{ik}$ is now defined as a cone with a vertex at the center of the $i^{\th}$ voxel and an angular range given by $\psi_k(\s)$.
In our analysis, we consider a path between voxels with indices $i_0$, $i_1$, \ldots $i_n$.
This path can be considered to consist of $n$ subpaths between $i_0$ to $i_1$, $i_1$ to $i_2$ and so on until $i_{n-1}$ to $i_n$.
A schematic describing the above notations and other notations is presented in Fig.~\ref{fig:notation_path}. 

The probability of a particular path $\P$, i.e.~$\Pr(\P | \emis, \att)$, is the ratio of the flux incident on the detector through the path $\P$ to the flux incident on the detector through all possible paths.
Let $\Phi(\P)$ denote the flux of photons incident on the detector through a path $\P$. Then,
\begin{equation}
\Pr(\P | \emis, \att) = \frac{\Phi(\P)}{\sum_{\P'} \Phi(\P')}.
\label{eq:exp_pr_path}
\end{equation}
To derive the expression for $\Phi(\P)$, we first define a few other radiometric quantities. 
Let $\Q$ denote the radiant energy, the 3D vector $\r$ denote a location in space and $\E$ denote energy. 
The fundamental radiometric quantity we use to describe the photon transport is the photon distribution function $w(\r, \s, \E)$, given by
\begin{equation}
w(\r, \s, \E) = \frac{1}{\E} \frac{\partial^3 \Q}{\partial V \partial \Omega \partial \E}.
\end{equation}
%In terms of photons, $w(\r, \s, \E, t)$ describes the number of photons contained in volume $\Delta V$ centered on the 3D vector $\r$ and traveling in solid angle $\Delta \Omega$ about direction $\s$ with energies between $\E$ and $\E + \Delta \E$.
%The relation between the photon distribution function and flux is given by
%\begin{equation}
%\Phi = c_m \int d^3 r \int d^2 \s \int d \E \ (\n \cdot \s) w(\r, \s, \E) 
%\end{equation}
The quantity used to describe emission of photons is the source distribution function, denoted by $\Xi(\r, \s, \E)$, and defined as
\begin{equation}
\Xi (\r, \s, \E) = \frac{\partial^3 \Phi}{\partial V \partial \Omega \partial \E}.
\end{equation}  
%This quantity can be interpreted as a number of photons injected per second into volume $\Delta V$ in an energy range $\Delta \E$ over solid angle $\Delta \Omega$.
Finally, the radiant intensity, denoted by $\Gamma(\s)$ is defined as
\begin{equation}
\Gamma(\s) = c_m \int  \int  w(\r, \s, \E)d^3 \r d \E.
\label{eq:def_radint}
\end{equation}

Consider a point source at location $\rs$ in the voxel indexed by $i_0$ emitting photons of energy $\E_0$ at a constant rate $\lambda_{i_0}$. 
Assuming isotropic emission, the source distribution along the sub-path $\subpath_{i_0, k_0}$ is given by
\begin{equation}
\Xi(\r, \s, \E) = \frac{\lambda_{i_0}}{4 \pi} \delta( \r - \rs) \delta ( \E - \E_0) \psi_{k_0} (\s),
\label{eq:src_dist_fn}
\end{equation}
where $\delta(x)$ is the Dirac delta function. 
As photons travel from the voxel $i_0$ to $i_1$, a fraction of the photons scatter, leading to attenuation of photons along this path.
The effect of the attenuation transform on the distribution function $w$ is given by \cite{barrett2013foundations}
\begin{equation}
[\X \w](\r, \s, \E) = \frac {1}{c_m} \int_0^{\infty}   w(\r - \s l) \exp \left[- \int_{0}^l  \mu (\r - \s l', \E)d l' \right] d l, 
\end{equation} 
where $\mu (\r, \E)$ denotes the attenuation coefficient at location $\r$ and energy $\E$, and $c_m$ denotes the speed of light.
Applying the attenuation transform to the source distribution function (Eq.~\eqref{eq:src_dist_fn}) yields 
\begin{align}
[\X \Xi] (\r, \s, \E) = \frac{\lambda_{i_0}}{4 \pi \cm} \delta ( E - E_0)\psi_{k_0} (\s) \int_{0}^{\infty}  \delta ( \r - \rs - \s l) \exp  \left[ -\int_{0}^{l}  \mu ( \r - \s l',\E)d l'  \right]d l .  
\label{attn_oper}
\end{align}
Now, it can be shown that \cite{barrett2013foundations}
\begin{equation}
\int_{l=0}^{\infty}  \delta ( \r - \rs - \s l) d l = \frac{1}{|\r - \rs |^2 } \delta \left( \s - \s_{10} \right),
\end{equation}
where
\begin{align}
\s_{10} = \frac{ \r - \rs} { |  \r - \rs |}.
\label{def_s10}
\end{align}
Using the above relation and the sifting property of the delta function, the integral over $l$ in Eq.~\eqref{attn_oper} can be simplified to yield
\begin{align}
[\X \Xi](\r, \s, \E) &= \frac{\lambda_{i_0}}{4 \pi \cm | \r - \rs |^2 } \exp \left[- \int_{0}^{| \r - \rs |}   \mu ( \r - \s l',\E)d l' \right] \times \nonumber\\
&\delta \left( \s - \s_{10}  \right) \delta( \E - \E_0) \psi_{k_0} (\s).
\label{attn_oper_3}
\end{align}
In the considered path, Compton scattering occurs at some location $\r$ within voxel $i_1$.
This operation can be described in terms of the scattering operator $\K$ as
\begin{equation}
[\K w] ( \r, \s, \E) = \int  \int_{4 \pi }  K ( \s, \s', \E, \E' | \r) w ( \r, \s', \E')d \Omega'd \E',
\label{scat_oper_1}
\end{equation}
where $K ( \s, \s', \E, \E' | \r)$ denotes the scattering kernel and is given by
\begin{align}
K( \s, \s', \E, \E' | \r) = \cm n_{\sc}(\r) \frac{\partial \sigma_{sc}}{\partial \Omega} \delta \left\{ \E - \left[ \frac{1}{\E'} + \frac{1}{m c^2} ( 1 - \cos \theta) \right]^{-1} \right\},
\label{scat_kernel}
\end{align}
where $n_{\sc} (\r)$ is the density of scatterers and is related to the scattering coefficient by
\begin{equation}
\mu(\r) = n_{\sc}(\r) \sigma_{\sc},
\label{musc_nsc}
\end{equation}
where $\sigma_{\sc}$ is the scattering cross section. 
Also, $\theta$ is the angle at which the outgoing photon scatters relative to the incoming photon, so $\cos \theta = \s \cdot \s'$.
%For the considered path, we determine photons that scatter and travel towards spatial coordinate $\r'$ in voxel $i_2$.
%In this case, the value of $ \theta$, which we denote by $\theta_{210}$ is 
%\begin{equation}
%\cos \theta_{210} = \left( \s_{21}\right) \cdot \left( \s_{10} \right),
%\end{equation}
%where 
%\begin{align}
%\s_{21} = \frac{ \r' - \r} { |  \r' - \r |},
%\end{align}
%and $\s_{10}$ is as defined by Eq.~\eqref{def_s10}.
Finally, the differential scattering cross section $\frac{\partial \sigma_{sc}}{\partial \Omega}$ in Eq.~\eqref{scat_kernel} is given by the Klein-Nishina formula \cite{Zaidi_2003}.
For notational simplicity, define $K_{\magni}(\s, \s', \E| \r)$ by 
\begin{equation}
K_{\magni}(\s, \s', \E| \r) =  n_{\sc} (\r) \frac{\partial \sigma_{sc}}{\partial \Omega} \bigg|_{\cos \theta = \s \cdot \s' }.
\end{equation}
Also, denote the path integral between any two locations $\r_l$ and $\r_m$ by the function $\gamma(\r_{l}, \r_{m},\E)$, i.e.
\begin{equation}
\gamma(\r_{l}, \r_{m},\E) =  \int_{0}^{ | \r_{l} - \r_{m} | }   \mu \left( \r_{l} - t \frac {\r_{1} - \r_{m}}{ | \r_{1} - \r_{m} |}, \E\right)\dr t.
\label{eq:pathintegral}
\end{equation}
Substituting the expression for the {attenuation transformed source} distribution function from Eq.~\eqref{attn_oper_3} into Eq.~\eqref{scat_oper_1}, and using the sifting property of the delta function in angular coordinates yields
\begin{align}
[\K \X \Xi](\r , \s, \E ) &= \frac{\lambda_{i_0}}{4 \pi \cm | \r - \rs |^2 } K\left(\s, \s_{10},\E, \E_0|\r\right) \exp \left[- \gamma(\r, \rs, \E_0) \right] \psi_{k_0} (\s_{10}).
\label{eq:KXE_w}
\end{align}
We now integrate this distribution function over all possible locations within the $i_1^{\th}$ voxel and over all possible energies.
This yields the radiant intensity along direction $\s$ due to photons traveling through the $\subpath_{i_0,k_0}$ subpath and scattering within the $i_1^{\th}$ voxel. 
%Integrating over all possible locations within the $i_1^{\th}$ voxel that lie in the sub-path $\subpath_{i_0,k_0}$, and over all possible energies yields the radiant intensity (Eq.~\eqref{eq:def_radint}) due to the radiation in $\subpath_{i_0,k_0}$ subpath that scatters in the $i_1^{\th}$ voxel along direction $\s$.
Denote this radiant intensity by $\Gamma_{i_0 i_1 k_0}(\s)$.
Substituting the distribution function from Eq.~\eqref{eq:KXE_w} into Eq.~\eqref{eq:def_radint} yields
\begin{align}
& \Gamma_{i_0 i_1 k_0} (\s) =  \int  \frac{\lambda_{i_0}}{4 \pi | \r - \rs |^2 }  K_{\magni} \left(\s, \s_{10},\E_0|\r\right)  \exp \left[- \gamma(\r, \rs, \E_0)  \right] \phi_{i_1}(\r)\psi_{k_0} (\s_{10})d^3 \r , 
\label{Nph_after_scat}
\end{align}
where $\phi_i(\r)$ is the voxel basis function defined as below:
\begin{equation}
\phi_i(\r) = \begin{cases}
1, \quad\r \ \text{lies within voxel i.} \\
0, \quad\mathrm{otherwise.}
\end{cases} 
\end{equation}
Assuming that the functions $K\left(\s, \s_{10},\E, \E_0\right)$ and $\exp \left[- \gamma (\r, \rs, \E_0) \right]$  do not vary relatively within any location within voxel $i_1$, we can evaluate them when $\r$ is the center of the $i_1^{\th}$ voxel and $\rs$ is the center of the $i_0^{\th}$ voxel.
Denoting the center of the $i_0$ and $i_1$ voxels by $\r_{0}$ and $\r_{1}$,  respectively, and denoting the direction vector joining $\r_{0}$ and $\r_{1}$ by $\s_{c10}$, we obtain
\begin{align}
& \Gamma_{i_0 i_1 k_0}(\s) \approx \frac{\lambda_{i_0}}{4 \pi }\exp [- \gamma(\r_1, \r_0, \E_0)] K_{\magni}\left(\s, \s_{c10},\E_0|\r_1\right) \int  \frac{1}{| \r - \rs |^2}\psi_{k_0} (\s_{10}) \phi_{i_1}(\r)d^3 \r .
\end{align}
Using the definition of $\s_{10}$ from Eq.~\eqref{def_s10} and denoting $|\r - \rs|$ by $R$, we perform a change of variables $\r - \r_s = R \s_{10}$ and replace $\r$ by $ R^2 d R d \s_{10}$.
Simplifying further yields 
\begin{align}
 \Gamma_{i_0 i_1 k_0}(\s)  \approx &  \frac{\lambda_{i_0}}{4 \pi }\exp [- \gamma(\r_1, \r_0, \E_0)] \times\nonumber\\
& K_{\magni}\left(\s, \s_{c10},\E_0|\r_1\right)  \int   \psi_{k_0} (\s_{10}) \int  \phi_{i_1}(\r_s + R \s_{10})d R ~d \s_{10}.
\label{Nph_after_scat_2}
\end{align}
The integral over $R$ is equal to the distance traversed by the $\r_s + R \s_{10}$ vector within the $k_1^{\th}$ voxel, so this distance should vary with $\s_{10}$. 
However, assuming that this variation is not substantial, we approximate it by the distance that is covered by the vector $\r_0 + R \s_{k_0}$ in the $i_1^{\th}$ voxel, which we denote by $\Delta_{i_1}(\subpath_{i_0 k_0})$. Note that $\s_{k_0}$ is the discretized direction vector associated with the $k_0^{\th}$ ordinate.
Further, performing the integral over $\s_{10}$ yields
% $\psi_{k_0}(\s)$ is unity over a range $\Delta \Omega$.
the following expression for $\Gamma_{i_0 i_1 k_0}(\s)$:
\begin{align}
\Gamma_{i_0 i_1 k_0}(\s) \approx \frac{\lambda_{i_0}}{4 \pi }\exp [-\gamma(\r_1, \r_0, \E_0)]  K_{\magni}\left(\s, \s_{c10},\E_0|\r_1\right) \Delta_{i_1}(\subpath_{i_0 k_0})\Delta \Omega.
\label{eq:del_subpath}
\end{align} 

%where $\Delta \Omega$ is the solid angle subtended by the sub-path. 
To proceed further, for mathematical tractability, assume that this entire radiant intensity is concentrated at the center of the $i_1^{\th}$ voxel, i.e.~at location $\r_1$ and divided uniformly over the subpath from this voxel that includes the direction $\s$.
Denote this subpath by $\subpath_{i_1, k_1}$. 
Thus, the distribution function along this subpath, denoted by $w_1(\r, \s, \E)$, is given by
\begin{align}
& w_1(\r , \s, \E ) =  \frac{\lambda_{i_0}}{4 \pi \cm } \exp \left[- \gamma(\r_1, \r_0, \E_0) \right] K\left(\s, \s_{c10},\E, \E_0|\r_1\right) \Delta_{i_1}(\subpath_{i_0 k_0}) \delta(\r - \r_{1}) \psi_{k_1}(\s).
\end{align}
After this scattering operation, the photons along this path suffer from attenuation as they travel towards voxel $i_2$. 
%This attenuation operation leads to the photon distribution function at a location $\r$ inside the $i_2^{\th}$ voxel as
%\begin{align}
%[\X w_1] (\r, \s, \E) = \frac{\lambda_{i_0} \Delta_{i_1}(\subpath_{i_0 k_0})}{4 \pi \cm^2 |\r - \r_1|^2 } \exp \left\{ - \gamma(\r, \r_1) - \gamma(\r_1, \r_0) \right \}  K\left( \s_{21}, \s_{10}, E_0, E_1 \right)  \psi_{k_1}(\s) \delta (\s - \s_{21}).
%\end{align} 

This series of operations continues until the path intersects with the detector.
In the considered path, the last voxel where scattering occurs is $i_n$ and the subpath that connects the last voxel to the detector is denoted by $\subpath_{i_n, k_n}$.
Denote the coordinates on the front face of the detector by $\r_d$, and the normal unit vector to the detector plane by $\n$.
Then the distribution function at the face of the detector, denoted by $w_d(\r_d, \s, \E)$ is given by
\begin{align}
w_d(\r_d, \s, \E) &= \frac{ \prod_{m=0}^{n-1} \Delta_{i_{m+1}}(\subpath_{i_m k_m})}{ |\r_d - \r_{n}|^2 }  \exp \{ - \gamma(\r_1, \r_0, \E_0)  - \ldots \gamma(\r_{d}, \r_n, \E_n) \}  \times \nonumber \\
& \prod_{m=0}^{n-1} \{ K_{\magni}\left(\s_{c,m+2,m+1}, \s_{c,m+1,m},  \E_{m} |\r_{m+1}\right)\}\delta( \E - \E_n) \psi_{k_n} (\s) .
\end{align} 
%\hlc[cyan]{where $\s_{c,n+1,n}=\s_{k_n}$. Being consistent with the notation, $\s_{k_n}$ denotes the discretized direction vector corresponding to the $k_n^{\th}$ ordinate.}
Now the plane of the front face of the detector is given by $\delta(p - \r_d \cdot \n)$, where $p$ is the perpendicular distance from the origin to the detector plane.
Let the sensitivity of the collimator to a photon emitted from location $\r$ in direction $\s$ be denoted by $t(\r, \s)$.
Then, the flux detected through the considered path is given by
\begin{align}
& \Phi(\path) = \cm \int  \int  \int  t( \r_n, \s) (\n \cdot \s ) \delta( p - \r_d \cdot \n) w_d(\r_d, \s, \E)~d^3 \r_d ~d \Omega ~d E \nonumber \\
& = \frac{\lambda_{i_0}}{4 \pi} \int \int_{2 \pi} \int t( \r_n, \s) (\n \cdot \s ) \delta(p- \r_d \cdot \n)  \times \nonumber \\
& \exp \{ - \gamma(\r_1, \r_0, \E_0) - \ldots \gamma(\r_d , \r_n, \E_n) \} \frac{ \prod_{m=0}^{n-1} \Delta_{i_{m+1}}(\subpath_{i_m k_m})}{ |\r_d - \r_{n}|^2 } \times \nonumber \\
& \prod_{m=0}^{n-1} \{ K_{\magni}\left(\s_{c,m+2,m+1}, \s_{c,m+1,m}, \E_{m}|\r_{m+1} \right) \}  \delta( \E - \E_n) \psi_{k_n} (\s)~d^3 \r_d ~d \Omega ~d E .
\label{path_flux_1}
\end{align} 
Perform a change of variables by replacing $\r_d - \r_n$ by $R \s$, so that $d^3 r_d = R^2 d R d \s $.
Next, integrating over $E$ and $\Omega$ using the sifting property of the delta function yields 
\begin{align}
& \Phi(\path) = \frac{ \lambda_{i_0}}{4 \pi} \int  \int  \delta(p- \r_n \cdot \n - R \s \cdot \n) t\left(\r_n, \s \right)  (\n \cdot \s) \times \nonumber \\ 
& \exp \{ - \gamma(\r_1, \r_0, \E_0) - \ldots \gamma(\r_n + R \s , \r_n, \E_n) \} \prod_{m=0}^{n-1} \{ \Delta_{i_{m+1}}(\subpath_{i_m k_m}) \} \times  \nonumber \\
&  \prod_{m=0}^{n-1}  \{ K_{\magni} \left(\s_{c,m+2,m+1}, \s_{c,m+1,m}, \E_{m} |\r_{m+1}\right) \} \psi_{k_n} (\s) ~d \s ~d R.
\label{path_flux_2}
\end{align}

The above expression formalizes the radiation through a path for a general SPECT system. 
We now derive the specific form of this expression for a SPECT system with a parallel-hole collimator.
This collimator allows only photons that are incident in a small range of angles around $\n$ to pass through the collimator. 
Thus, assuming $\n \cdot \s \approx 1$ when $t(\r, \s) > 0$, the integral over $R$ can be performed to yield
\begin{align}
& \Phi(\path) = \frac{ \lambda_{i_0}}{4 \pi} \int   t\left(\r_n, \s \right)  \exp \{ - \gamma(\r_1, \r_0, \E_0) - \ldots \gamma(\r_{d_0} , \r_n, \E_n) \} \times  \nonumber \\
& \prod_{m=0}^{n-1}   \{ \Delta_{i_{m+1}}(\subpath_{i_m k_m}) \}\prod_{m=0}^{n-1}  \{ K_{\magni} \left(\s_{c,m+2,m+1}, \s_{c,m+1,m},  \E_{m}|\r_{m+1} \right) \} \psi_{k_n} (\s) ~d \s,
\label{path_flux_3}
\end{align}
where $\r_{d0} = \r_n + (p - \r_n \cdot \n) \s$ is the coordinate on the detector plane where the gamma-ray photon is incident. 
Further, assuming the angular ordinates in the object space to be small compared to the angular range allowed by the collimator, we replace $\s$ with $\s_{k_n}$, where $\s_{k_n}$ denotes the discretized direction vector corresponding to the $k_n^{\th}$ ordinate.. Then the integration over $\s$ can be performed to yield 
\begin{align}
& \Phi(\path) = \frac{ \lambda_{i_0} \Delta \Omega}{4 \pi}  \exp \{ - \gamma(\r_0, \r_1, \E_0)  - \ldots \gamma(\r_{dc0} , \r_n, \E_n) \} t\left(\r_n, \s_{k_n} \right) \nonumber \\
&  \prod_{m=0}^{n-1} \Delta_{i_{m+1}}(\subpath_{i_m k_m}) \prod_{m=0}^{n-1} K_{\magni}\left(\s_{c,m+2,m+1}, \s_{c,m+1,m}, \E_{m}|\r_{m+1} \right),
\label{path_flux_4}
\end{align}
where $\r_{dc0} = \r_n + (p - \r_n \cdot \n) \s_{k_n}$.
To simplify the expression for $\Phi(\path)$, note that in our analysis, we need to only consider the dependence of this expression on the emission and attenuation coefficients.
Quantities that are not dependent on these parameters in the above expression are denoted by $\Lambda(\path)$. 
Further, to simplify notation, we define $s_{\eff}(\P)$ as
\begin{equation}
s_{\eff}(\P) = \Lambda(\path) \exp \left[ -\sum_{m=0}^{n} \gamma(\r_m, \r_{m+1}, \att(\E_m)) \right]  \prod_{m=1}^{n} \mu_{k_m}(E_{m-1}).
\label{seff_path}
\end{equation}
This term actually denotes the effective sensitivity for path $\path$ to the detector. 
Further, denote the activity in the starting voxel of the path, i.e. $\lambda_{i_0}$, by $\lambda(\path)$.
Eq.~\eqref{path_flux_4} can then be rewritten as 
\begin{equation}
\Phi(\P) = s_{\eff}(\P) \lambda(\P).
\end{equation}
The attenuation coefficient in Eq.~\eqref{seff_path} is a function of energy, which does not lend itself to Fisher information matrix analysis. To address this issue, based on the energy spectrum for common radionuclides such as Technetium-99m (Tc99m), we model the energy dependence of the attenuation coefficent  \cite{xcom3} as a linear function of energy, i.e.
\begin{equation}
\att(\E_m)= a\att+b.
\end{equation}
Here $\attn$ denotes the attenuation coefficient at the photo-peak energy i.e. $\attn=\attn(E_0)$.
Also, $a$ and $b$ denote constants that can be computed from the spectrum. Finally, substituting the expression of $\Phi(\P)$ in Eq.~\eqref{eq:exp_pr_path} yields:
\begin{equation}
\Pr(\path | \emis, \att) = \frac{\lambda(\path) s_\eff(\P)}{\sum_{\pathp} \lambda(\pathp) s_{\eff}(\pathp)}.
\label{eq:pr_path_2}
\end{equation}
This explicit modeling of probability of path in terms of $\emis$ and $\att$ is then used to express the log-likelihood and Fisher information analysis.
%We use Monte Carlo integration to evaluate these expressions using simulated LM data.

\subsection{Computing the PDF $\pr(\mathbf{\hat{A}}_j | \path)$}
\label{sec:pr_Ajhat_path}
The term $\pr(\bAhat_j | \P)$ denotes the PDF of the measured attribute vector $\bAhat_j$ given the photon followed a particular path $\path$.
This attribute vector, as mentioned above, consists of the position of interaction of the gamma-ray photon with the crystal, denoted by $\hat{\r}_j$, and the energy deposited by the photon in the crystal, denoted by $\hat{\E}_j$.
%Due to the finite energy and spatial resolution of detectors and due to the uncertainty in the estimation process, the estimated attribute vector $\bAhat_j$ differs from the true attribute $\bA_j$. 
%To model these sources of randomness, we first write $\pr(\bAhat_j | \path)$ as
To model the randomness in estimating $\bAhat_j$ due to the uncertainty in the estimation process and the finite energy and spatial resolution of the detectors, we first write $\pr(\bAhat_j | \path)$ as
\begin{equation}
\pr(\bAhat_j | \path) = \int \pr(\bAhat_j | \bA_j) \pr(\bA_j  | \path ) d \bA_j. 
\label{pr_Ajhat}
\end{equation}
To obtain the expression for $\pr(\bA_j| \path)$, consider a path that connects several voxels, as in the section above.
Consider a photon that propagates exactly between the center of these voxels before reaching the detector.
Denote the energy of the photon at the end of the path by $E_{\path}$ and the location where the photon interacts with the detector by $\r_\path$.
Assuming that the paths are discretized finely enough, we can assume that all the photons within this path will have approximately the same energy and interact with the detector at the same location. 
Thus, we can write 
\begin{equation}
\pr(\bA_j | \path)  = \pr(\r, \E  | \path) \approx \delta (\bA_j - \bA_j({\path})),
\label{pr_Aj}
\end{equation}
where $\bA_j({\path})=\{\r_{\path},\E_{\path}\}$.

Next, to derive the expression for the term $\pr(\bAhat_j | \bA_j)$, assume that the finite spatial and energy resolution of the detector and the uncertainty in the estimation of the LM attributes can be modeled by a Gaussian distribution.
This is a very reasonable assumption for most SPECT systems \cite{Kojima:89, Jha:15:pmb}. 
%The variance of this distribution is given by the intrinsic spatial and energy resolution of the detector.
%Denote the intrinsic position resolution in the $x$, $y$ and $z$ axis by $\sigma_x$, $\sigma_y$ and $\sigma_z$ and the energy resolution by $\sigma_E$.
%Define a diagonal covariance matrix $\Kdet$ with the diagonal entries as $\sigma_x$, $\sigma_y$, $\sigma_z$ and $\sigma_E$.
Then, we can write
\begin{equation}
\pr(\bAhat_j | \bA_j) = \frac{1}{ (2 \pi)^2 |\Kdet|} \exp \left[ -\frac{(\bAhat_j - \bA_j)^{\dagger} \Kdet^{-1} (\bAhat_j - \bA_j)}{2}\right],
\label{pr_Ajhat_Aj}
\end{equation} 
where $\Kdet$ denotes the covariance matrix quantifying the variances and co-variances in the energy and position estimates, and where $|\Kdet|$ denotes the determinant of the matrix $\Kdet$.
Substituting Eq.~\eqref{pr_Ajhat_Aj} and~\eqref{pr_Aj}  into Eq.~\eqref{pr_Ajhat}, and using the sifting property of the delta function yields 
\begin{equation}
\pr(\bAhat_j | \path) = \frac{1}{(2 \pi)^2 |\Kdet| } \exp\left[-\frac{ (\bAhat_j - \bA_j(\path))^{\dagger} \Kdet^{-1} ( \bAhat_j - \bA_j(\path)) }{2}\right].
\label{eq:pr_Ajhat_path}
\end{equation}
We now have the expressions for $\Pr(\path|\emis,\att)$ (Eq.~\eqref{eq:pr_path_2}) and $\pr(\bAhat_j | \path)$ (Eq.~\eqref{eq:pr_Ajhat_path}) as required to formalize the likelihood of the LM data. We now proceed to deriving the expression for the elements of the Fisher information matrix.
\subsection{The Fisher information matrix for the LM data}
\label{sec:Fisher_info}
The general expression for the elements of a FIM is given by
\begin{equation}
F_{q q'} = - \left\langle \frac{\partial^2 \L ( \emis, \att | \hat{\A}, T)}{\partial \theta_q \partial \theta_{q'}} \right\rangle_{(\Ahat, T| \emis, \att)},
\label{fim_general}
\end{equation}
where $\theta_q$ and $\theta_{q'}$ denote the parameters we intend to estimate, and thus in our case are the activity-attenuation coefficients in the $q^{\th}$ and $q'^{\th}$ voxels of the object, and where $\L( \emis, \att | \hat{\A}, T)$ is the log-likelihood of the observed LM data (Eq.~\eqref{loglikelihood2}).
Substituting the expression for $\Pr(\path| \emis, \att)$ from Eq.~\eqref{eq:pr_path_2} into Eq.~\eqref{loglikelihood2}, and further using Eq.~\eqref{prob_time} yields
\begin{align}
\L ( \emis, \att | \hat{\A}, T) = \sum_{j=1}^J \log \left[ \sum \pr (\bAhat_j | \path) \lambda(\path) s_{\eff} (\P) \right]  + (J-1) \log T - \beta T - \log (J-1)!.
\label{loglikelihood3}
\end{align}
Now, note that $\beta$, which is the mean rate of photons detected, is equivalently the total radiant flux over all possible paths. Thus, $\beta$ is given by
\begin{equation}
\beta =  \sum_{\pathp} \lambda(\pathp) s_{\eff}(\pathp).
\label{def_beta}
\end{equation}
Using Eq.~\eqref{def_beta} to substitute for $\beta$, and differentiating the log-likelihood with respect to the activity in the $q^{\th}$ voxel $\lambda_q$ yields 
\begin{equation}
\frac{\partial \L}{\partial \lambda_q} = \sum_{j=1}^J \frac{\sum_{\P_q} \pr(\bAhat_j | \P) s_{\eff} (\P) }{\sum_{\P} \pr(\bAhat_j | \P) \lambda(\P) s_{\eff} (\P)} - T {\sum_{\P_q}s_{\eff}(\P)},
\label{diff_likelihood_lambda}
\end{equation}
where the summation in the numerator is only over the paths that start from voxel $q$, denoted by $\path_q$. 
Similarly, differentiating the log-likelihood (Eq.~\eqref{loglikelihood3} with respect to the attenuation coefficient in the $q^{\th}$ voxel, i.e.~$\mu_q$, yields
\begin{align}
 \frac{\partial \L}{\partial \mu_q} =  &\sum_{j=1}^J \frac{\sum_{\P} \pr(\bAhat_j | \P) \lambda(\path) s_{\eff} (\P)  \left[ -\Delta_q(\P) + \frac{\zeta_q(\path)}{\mu_q} \right]}{\sum_{\P} \pr(\bAhat_j | \P) \lambda(\path) s_{\eff} (\P)} \times \nonumber\\
&- T {\sum_{\P} \lambda(\path) s_{\eff}(\P)}\left[ -\Delta_q(\path) + \frac{\zeta_q(\path)}{\mu_q} \right],
\label{diff_likelihood_mu}
\end{align}
where $\zeta_q(\path)$ and $\Delta_q(\path)$ are the number of scatter events occurring in the $q^{\th}$ voxel in the considered path and the distance that the considered path covers in the $q^{\th}$ voxel, respectively. 
To derive the FIM elements, we must differentiate Eqs.~\eqref{diff_likelihood_lambda} and \eqref{diff_likelihood_mu} further with respect to the activity and attenuation coefficients in some other $q'{\th}$ voxel, and then average over the observed LM data .
The derivations are detailed in Appendix A, and the final expressions are as below:
\begin{align}
\left\langle\frac{\partial^2 \L}{\partial \mu_{q'} \partial \mu_q } \right\rangle_{(\Ahat, T| \emis, \att)} &= - \left\langle \sum_{j=1}^J \frac{ \left\{ \sum_{\P} \pr(\bAhat_j | \P) \lambda(\P)  s_{\eff} (\P)  \left[ -\Delta_q(\path) + \frac{\zeta_q(\path)}{\mu_q} \right] \right\}}{{ \sum_{\P} \pr(\bAhat_j | \P) \lambda(\P) s_{\eff} (\P) }} \times \right. \nonumber \\
& \quad \left. \frac{\left \{ \sum_{\P} \pr(\bAhat_j | \P) \lambda(\P)  s_{\eff} (\P)  \left[ -\Delta_{q'}(\path) + \frac{\zeta_{q'}(\path)}{\mu_{q'}} \right] \right \}}{ \sum_{\P} \pr(\bAhat_j | \P) \lambda(\P) s_{\eff} (\P) } \right\rangle_{(\Ahat, T| \emis, \att)}, \label{fim_attn} \\
\left\langle\frac{\partial^2 \L}{\partial \lambda_{q'} \partial \lambda_q } \right\rangle_{(\Ahat, T| \emis, \att)} &= - \left\langle \sum_{j=1}^J \frac{\{\sum_{\P_q} \pr(\bAhat_j | \P) s_{\eff} (\P) \} \{\sum_{\P_{q'}} \pr(\bAhat_j | \P) s_{\eff} (\P) \}}{\{\sum_{\P} \pr(\bAhat_j | \P) \lambda(\path) s_{\eff} (\P) \}^2 } \right\rangle_{(\Ahat, T| \emis, \att)},
\label{fim_act} \\
\left\langle \frac{\partial^2 \L}{\partial \mu_{q'} \partial \lambda_q } \right\rangle_{(\Ahat, T| \emis, \att)} &= - \left\langle\sum_{j=1}^J  \frac{\left\{\sum_{\P_q} \pr(\bAhat_j | \P) s_{\eff} (\P) \right\}}{\sum_{\P} \pr(\bAhat_j | \P) \lambda(\path) s_{\eff} (\P) } \times \right. \nonumber \\
& \quad \left. \frac{\left\{\sum_{\P} \pr(\bAhat_j | \P) \lambda(\P) s_{\eff} (\P) \left[-\Delta_{q'}(\P) + \frac{\zeta_{q'}(\P)}{\mu_{q'}} \right] \right\}}{\sum_{\P} \pr(\bAhat_j | \P) \lambda(\path) s_{\eff} (\P)} \right\rangle_{(\Ahat, T| \emis, \att)}, \label{fim_cross} \\
 \left\langle \frac{\partial^2 \L}{\partial \lambda_{q'} \partial \mu_q } \right\rangle_{(\Ahat, T| \emis, \att)}  &=  \left\langle \frac{\partial^2 \L}{\partial \mu_{q} \partial \lambda_{q'}} \right\rangle_{(\Ahat, T| \emis, \att)}. \label{fim_cross2}
\end{align}
Since we cannot simplify these expressions further, we use Monte Carlo integration to evaluate these expressions from simulated LM data, thus yielding the elements of the FIM. The inverse of the FIM yields the Cramer-Rao bound (CRB), which is the lower bound on the variance of any unbiased estimate of the activity and attenuation coefficients from the SPECT emission data. Thus, using the CRB, we can quantify the information content of the SPECT emission data on the task of jointly estimating activity and attenuation.
\section{Methods}
\label{sec:Implementation}

\subsection{SPECT imaging system and LM data acquisition}
\label{sec:sim_spect}
To evaluate the information content in the LM emission data for joint reconstruction, we simulated a clinical 2-D parallel-hole SPECT system configuration similar to GE's Optima NM/CT 640 SPECT/CT system. 
Details of this SPECT system geometry are given in Section \ref{sec:experiments}.
The geometric sensitivity of the detector to different paths, the finite extent and bore diameters of the collimator, and the finite energy and spatial resolution of the detectors were all modeled. 
For generating the LM data, we simulated the photon transport via a Monte Carlo-based approach in MATLAB. The scattering was modeled using the Klein-Nishina formula, which was normalized for a 2-D system so that all the scattering was in plane. While simulating the photon transport, we considered photons that scattered at most once or twice, depending on the experimental setup. 
This was done to reduce the computational requirements in the FIM computation code, as we discuss in the next section.
For each detected photon, the 1-D estimated position of interaction of the incident gamma-ray photon with the scintillator, the estimated energy of the gamma-ray photon, and the angular orientation of the detector were recorded in LM format.

\subsection{Implementing the Fisher information approach}
\label{sec:impl_fim}

For computing the FIM using the proposed path-based approach, software was developed in C programming language that used GPU acceleration using CUDA parallel computing platform.
The first step was to implement the path-based formalism to describe the radiation transfer through different paths, as described in Sec.~\ref{sec:rad_transfer_path}. 
This formalism was validated by comparing it with the results obtained using the Monte Carlo approach described 
above (Sec.~\ref{sec:sim_spect}) through several studies. The results obtained with the two approaches, for example, the photon flux, the energy spectrum, and also the profiles of the projection data, were found to match. 
We do not show these results since our focus in this paper is on studying the information content of LM data.

The validated path-based formalism was then used to develop software to compute the FIM terms, as described by Eqs.~\eqref{fim_attn}-\eqref{fim_cross2}.
For each detected LM event, we considered all the possible unscattered and scattered paths that the photon could have taken, regardless of the energy of the photon. 
In other words, photons were not labeled $\textit{a priori}$ as scattered or unscattered, since even a photon that falls in the photopeak (PP) window may have scattered.
As is convention, PP window was characterized as having a width equal to twice the full-width half maximum (FWHM) centered at the photopeak energy.
The FIM elements were computed at the true value of the parameters.
%The LM data simulated using the approach described above was used to compute the FIM.  
These FIM terms were used to compute the CRB, which was then used to compute the lower bound on the standard deviations for the activity and attenuation coefficients of the different voxels.

A major challenge in the FIM computation was the large computational and memory requirements. 
%The code scales as the number of voxels, the number of angular samples considered to define a path, the number of LM events, and finally the order of scatter events considered. 
These computational requirements were addressed using various algorithmic strategies. 
For example, to reduce the memory requirements, certain quantities that did not require large computation times were pre-computed and stored. 
This included quantities such as the radiological path between the different voxels, distance covered by a sub-path inside a voxel and sensitivity of the collimator as a function of the angular and spatial voxel index.
Further the code was parallelized and executed on graphics processing units (GPUs) for faster execution. More specifically, quantities that require sum over paths $\path$, such as $ \pr(\bAhat_j | \P) s_{\eff} (\P)$ and $\pr(\bAhat_j | \P) \lambda(\P)  s_{\eff} (\P)  \left[ -\Delta_q(\path) + \frac{\zeta_q(\path)}{\mu_q} \right]$, were computed in parallel and summed using reduction algorithm. 
A more detailed procedure to compute the FIM terms, including a pseudocode of the GPU-based implementation, are presented in Appendix B. 

However, even after these computational optimizations, we observed that the code took substantial time to execute. A major reason for this was that the code scales as the number of paths, $P$, which in turn scales as the square of the number of voxels and as the number of angular samples used to define a path.
 Further, if we consider photons that have scattered $M$ times, the number of path scales as $P^M$. To reduce this computational expenses and conduct experiments within a practical time limit, we considered phantoms with $32 \times 32$ pixels. Further, for most of the experiments, we modeled photons that scattered at most once; although in a subset of experiments, photons that scattered twice were modeled in the Fisher information computation code. 
In particular, in phantoms that had activity only within a single pixel, we could computationally model the dual-scattered photons in the Fisher information code within a practical time limit.
Thus, in summary, the code was implemented to model single and  dual-scattered events. 
\subsection{Experiments}
\label{sec:experiments}  
In the experiments, our objectives were to use the FIM computation framework to study the CRB of SPECT LM data for joint estimation of activity and attenuation distribution in simulation studies. 
The emission source was assumed to be Tc-99m, one of the most commonly used SPECT tracers, emitting photons at 140~$\mathrm{keV}$. %TODO
%The collimator and detector specifications were similar to those in a GE Discovery 760 system.
Photons were acquired at $64$ angular positions spaced uniformly over $360^{\circ}$. A low-energy high-resolution parallel-hole collimator, with specifications similar to GE Optima NM/CT 640 SPECT/CT scanner was simulated. The system yielded a resolution of 7.8~$\mm$ at 10~$\cmeter$ depth.
The scintillation detector had an intrinsic resolution of 4~$\mm$ and a length of 35~$\mm$. Further, the energy resolution of the scintillation detector was set to $10\%$ at $140~\keV$.

\begin{figure}
\centering
\subfloat[]{
\includegraphics[height = 1.5 in]{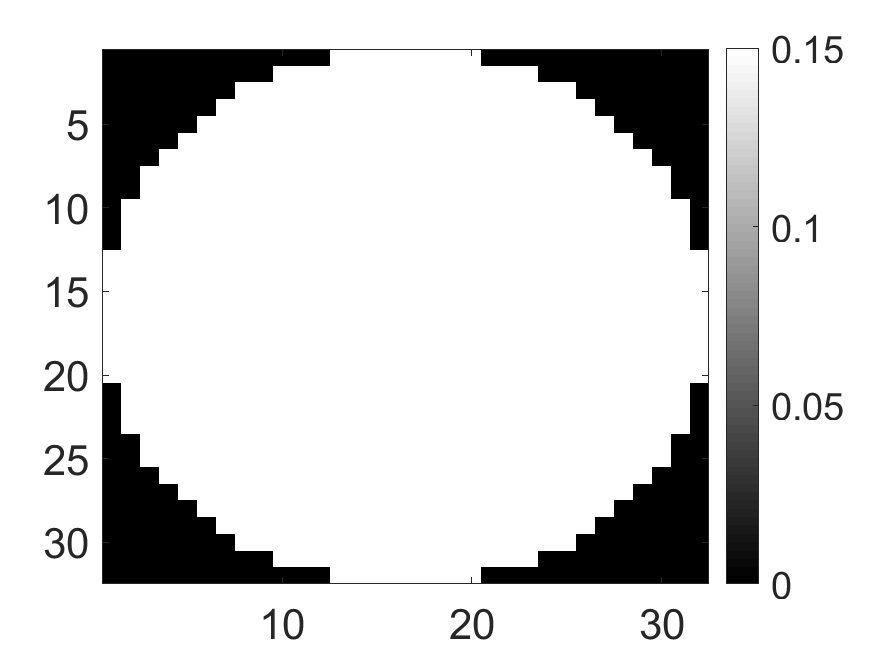}
\label{fig:uni_true}
}
\subfloat[]{
\includegraphics[height = 1.5 in]{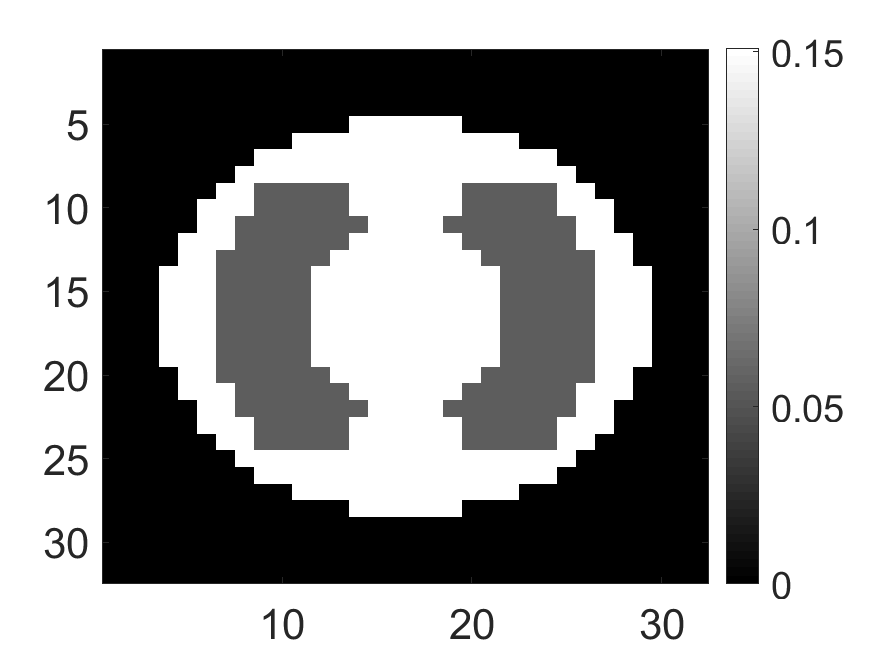}
\label{fig:nu_true}
}
\caption{True (a) uniform and (b) non-uniform attenuation map}
\label{fig:nu_uni_true}
\end{figure}

The system was used to first image a set of synthetic phantoms. For the synthetic phantoms, circular-shaped field of view with a diameter of $35~\cmeter$ and $32 \times 32$ spatial pixels were considered. Two attenuation map configurations were considered: a uniform attenuation map with a constant attenuation value of $0.15~\cmeter^{-1}$ inside the field of view (Fig.~\ref{fig:uni_true}) and a non-uniform attenuation map (Fig.~\ref{fig:nu_true}) that simulated the cardiac region with an attenuation coefficient of $0.151~\cmeter^{-1}$ and $0.056~\cmeter^{-1}$ in background and lung region, respectively. 
Three activity distributions were considered, namely phantom with radiotracer uptake confined to a single pixel (Fig.~\ref{fig:single_voxel_phantom}), uptake in multiple isolated pixels (Fig.~\ref{fig:multi_off_phantom}) and uptake over a donut-shaped region (Fig.~\ref{fig:donut_phantom}), referred to as single-pixel, multi-pixel and donut phantom, respectively. 
Experiments with these synthetic phantoms were conducted to gain an understanding of how activity distribution in uniform and non-uniform attenuation maps affect the information content of scattered photons.

Two anthropomorphic phantoms, namely the cardiac and the brain phantom, were also imaged to provide more clinical realism to our studies. The cardiac activity (Fig.~\ref{fig:xcat_act}) and attenuation (Fig.~\ref{fig:xcat_attn}) phantoms were generated using a 2D slice of extended cardiac and torso (XCAT) phantom \cite{segars20104d}. The brain activity (Fig.~\ref{fig:brain_act}) and attenuation (Fig.~\ref{fig:brain_attn}) phantoms were generated using a 2D slice of the Zubal phantom \cite{zubal1994computerized}.
For the study with the cardiac phantom, the SPECT system parameters were similar to as described above.
For the study with the brain phantom, we simulated a DaTScan SPECT study \cite{djang2012snm} by modeling a system that was imaging ioflupane (I-123) tracer emitting photons with energy of 159KeV and had a circular field-of-view with a diameter of $30~\cmeter$. 

LM data for these phantoms were generated using the simulated SPECT imaging system. In the first set of experiments, photons that scattered more than once were discarded.
The FIM for the LM data was computed and used to determine the CRB for the activity and attenuation estimates in the different pixels.
%Several experiments were conducted assuming that the source was confined to a single pixel, as shown in Figs.~\ref{fig:single_voxel_phantom} for easier interpretation of the results and for computational simplicity.
%Experiments with a more complicated donut-shaped activity uptake pattern, as shown in Fig.~\ref{fig:donut_phantom} were also conducted.
%Using the proposed approach, the CRB of the activity and the attenuation coefficients for all the pixels in a given phantom was computed from the entire LM SPECT emission data. 
%The CRB was used to compute the lower bound on the standard deviation of these coefficients.  
%The average of the standard deviation values for the attenuation coefficients in all the pixels with non-zero attenuation in a phantom yielded the mean standard deviation of the attenuation coefficient. 
%The standard deviation of the activity coefficient values was divided by the true activity coefficient in each pixel with non-zero activity and then averaged in all the pixels. 
%The mean standard deviation of the attenuation coefficients and the normalized mean standard deviation of the activity coefficients were used to quantitatively interpret the information content in LM data to estimate the attenuation and activity coefficients, respectively for the joint estimation task.  
%\hl{To assess the additional information that was provided from the scattered photons, we also computed the mean standard deviation values when using only the photopeak photons.}
The first experiment was conducted with $4 \times 10^5$ detected LM events, which is a clinically realistic count level in cardiac SPECT studies for a 2-D slice \cite{Frey}. Next, the detected photon counts were reduced from $4 \times 10^5$ to $40,000$ to study the CRB at different count levels. 
We also computed the CRB of the activity and attenuation coefficients when events only within the PP window were considered. This study was conducted to assess the impact of including the scattered photons on the CRB of these coefficients. 
Next, we studied the effect of varying the energy resolution of the system. 
%The energy resolution of the SPECT system was varied from 10$\%$ to 2.5$\%$ at $140~\mathrm{keV}$.  
%The CRB for the activity and attenuation coefficients was obtained for the phantom with activity in center voxel at different photon count levels.

In the second set of experiments, we considered photons that scattered at most twice. These experiments were conducted for the single-pixel activity phantom with non-uniform attenuation (Fig.~\ref{fig:nu_true}). This dual scattering was modeled in the FIM code. Our objective was to assess the impact of including dual-scattered photons on estimating the attenuation coefficient. In the first experiment, we varied the number of detected LM events and computed the CRB for attenuation coefficients. 
Our second experiment investigated the impact of change in attenuation coefficient on the CRB. The rationale behind this experiment was that an increase in the true value of the attenuation coefficient causes
an increase in the number of scattered photons, but also increases the proportion of dual-scattered photons. An important question is whether even in these scenarios, the scattered photons provide information to estimate the attenuation coefficients. 
In this experiment, we varied the attenuation coefficient in the lung region of the non-uniform phantom map (Fig.~\ref{fig:nu_true}). 

\section{Results}
\label{sec:Results}
\begin{figure}
\centering
\subfloat[]{
\includegraphics[height = 1.5 in]{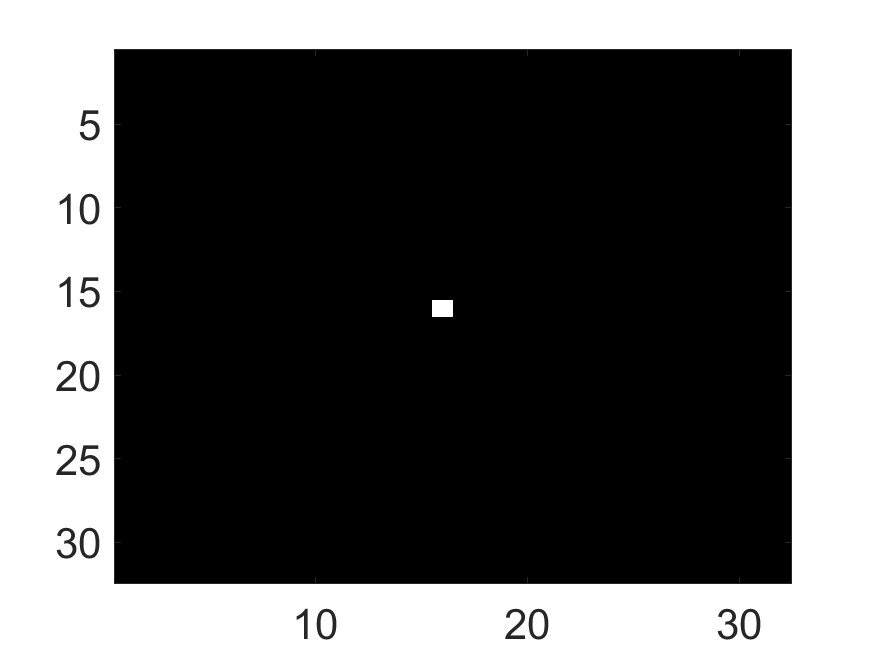}
\label{fig:single_voxel_phantom}
}
\subfloat[]{
\includegraphics[height = 1.5 in]{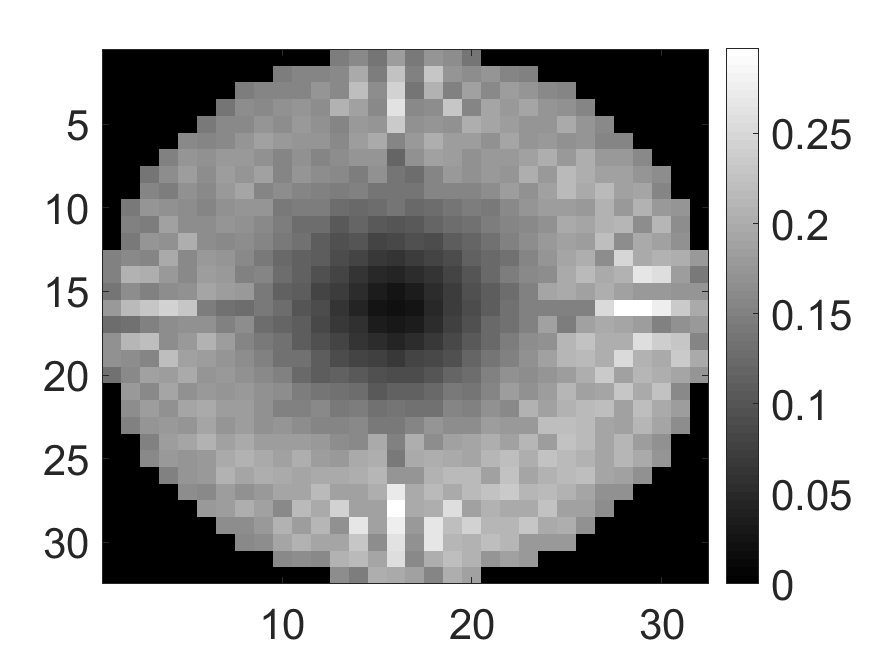}
\label{fig:stddev_single_vox}
}
\subfloat[]{
\includegraphics[height = 1.5 in]{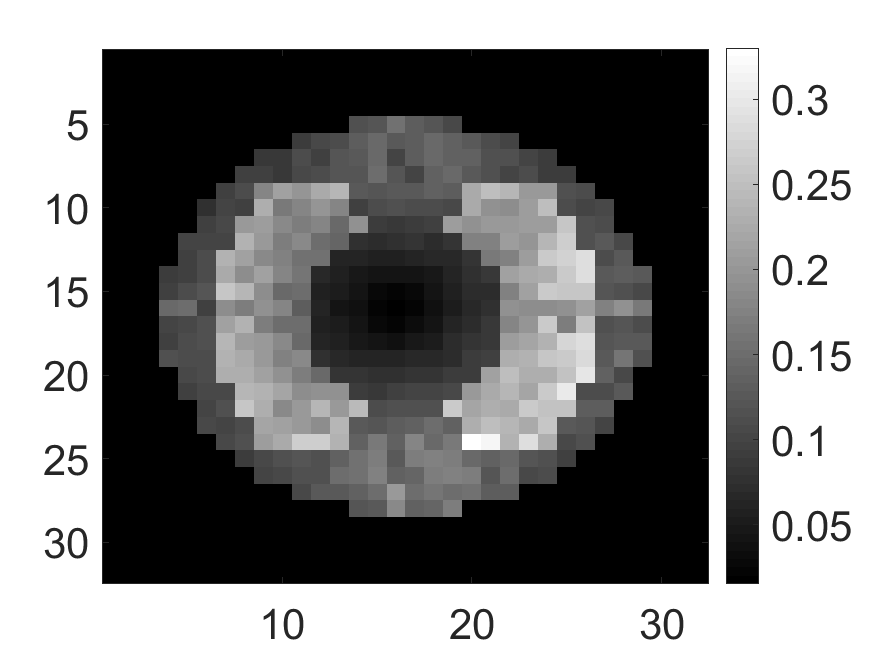}
\label{fig:stddev_single_vox_nu}
}
\hfill
\subfloat[]{
\includegraphics[height = 1.5 in]{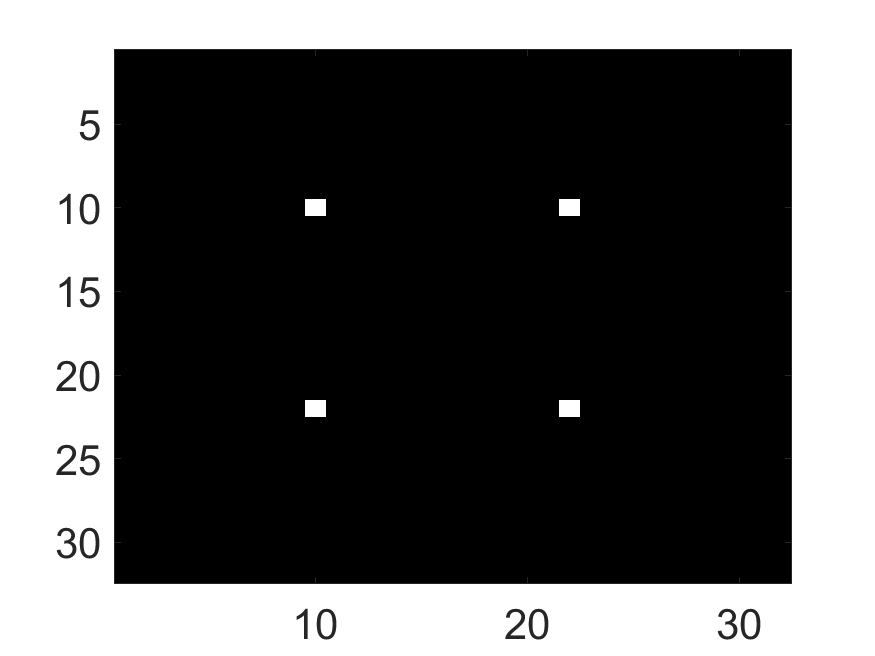}
\label{fig:multi_off_phantom}
}
\subfloat[]{
\includegraphics[height = 1.5 in]{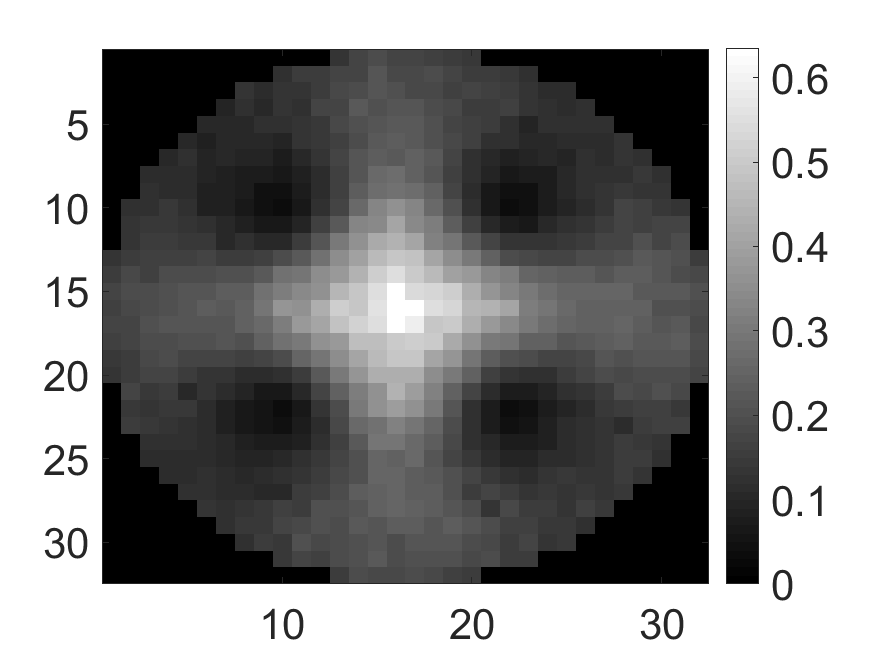}
\label{fig:stddev_multi_off}
}
\subfloat[]{
\includegraphics[height = 1.5 in]{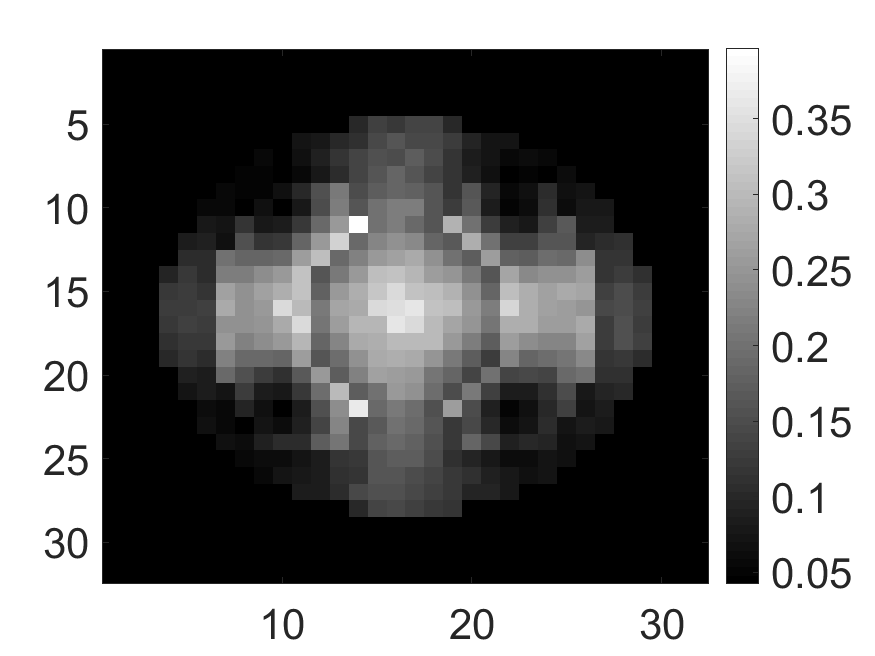}
\label{fig:stddev_multi_nu}
}
\hfill
\subfloat[]{
\includegraphics[height = 1.5 in]{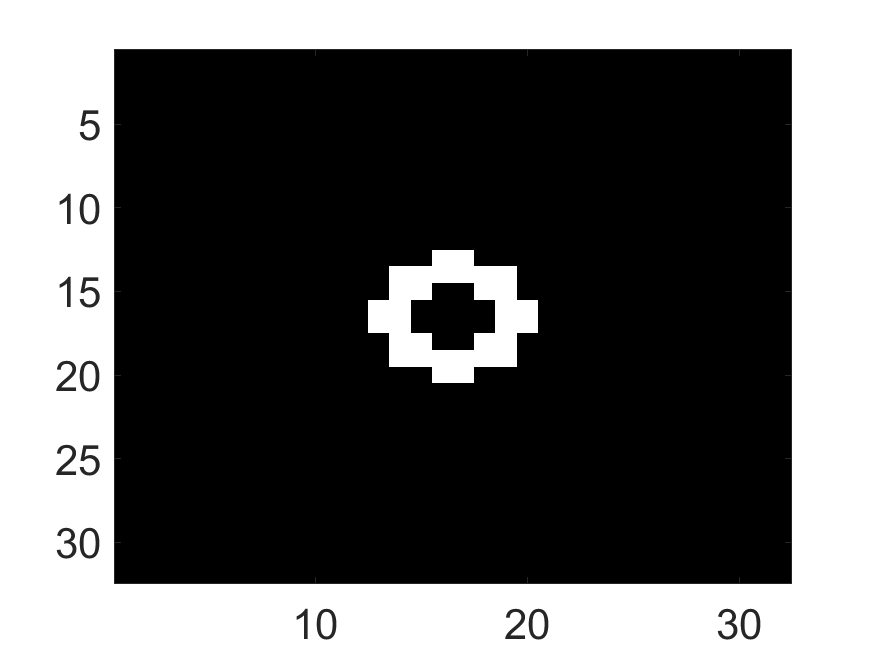}
\label{fig:donut_phantom}
}
\subfloat[]{
\includegraphics[height = 1.5 in]{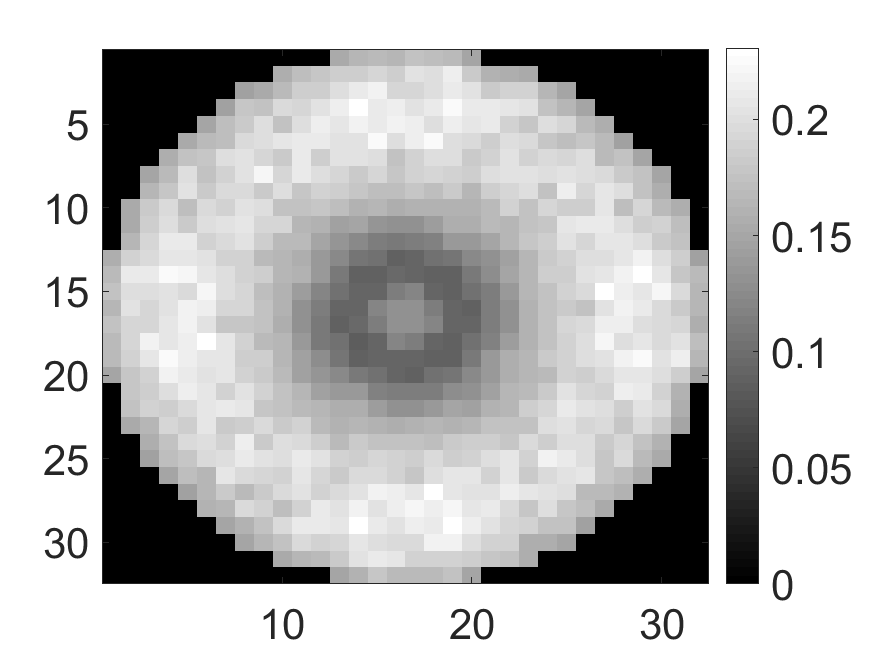}
\label{fig:stddev_donut}
}
\subfloat[]{
\includegraphics[height = 1.5 in]{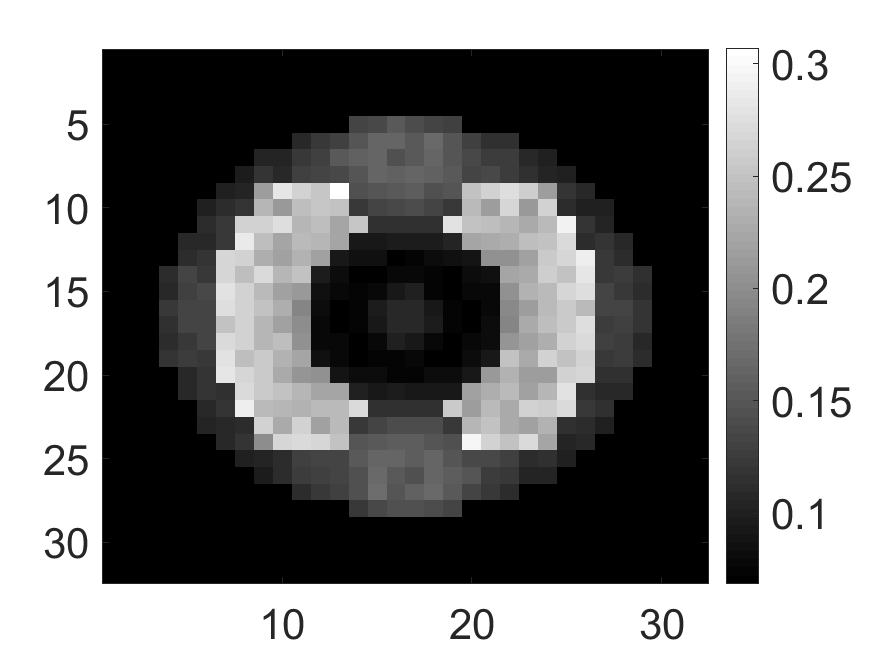}
\label{fig:stddev_donut_nu}
}
\caption{(a) The single-pixel phantom with on-axis activity. (d) The multi-pixel phantom with activity in off-axis locations. (g) The donut-shaped phantom. For uniform attenuation map, as shown in Fig.~\ref{fig:uni_true}, the normalized standard deviation of the estimate of the attenuation coefficients for the different pixels computed using the proposed approach for the (b) single-pixel phantom (e) multi-pixel phantom and (h) donut-shaped phantom. For non-uniform attenuation map (shown in Fig.~\ref{fig:nu_true}), the normalized standard deviation of the estimate of the attenuation coefficients for the different pixels computed using the proposed approach for the (c) single-pixel phantom (f) multi-pixel phantom and (i) donut-shaped phantom.}
\label{fig:stddev_phantoms}
\end{figure}

\begin{figure}
\centering
\subfloat[]{
\includegraphics[width = 2 in]{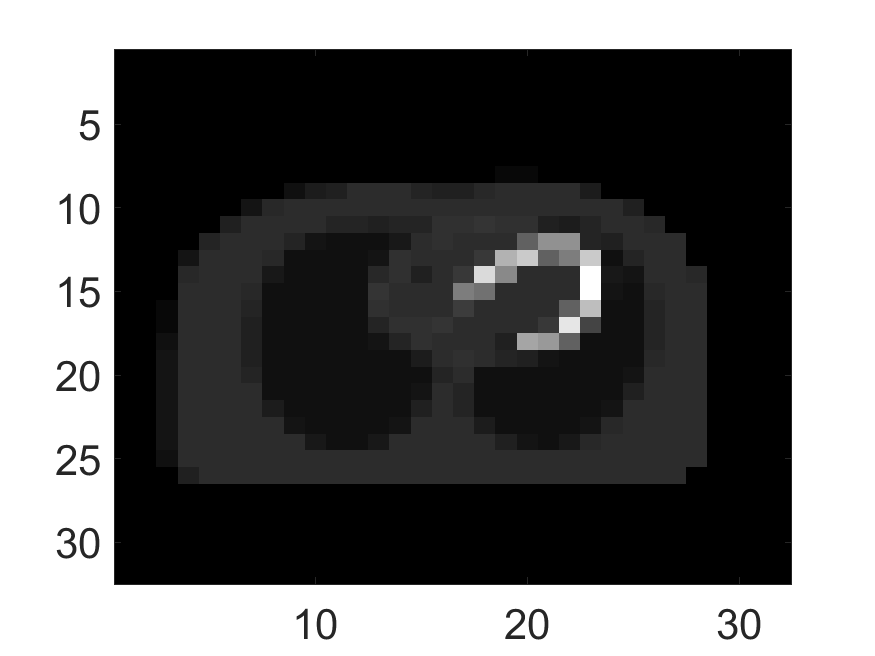}
\label{fig:xcat_act}
}
\subfloat[]{
\includegraphics[width = 2 in]{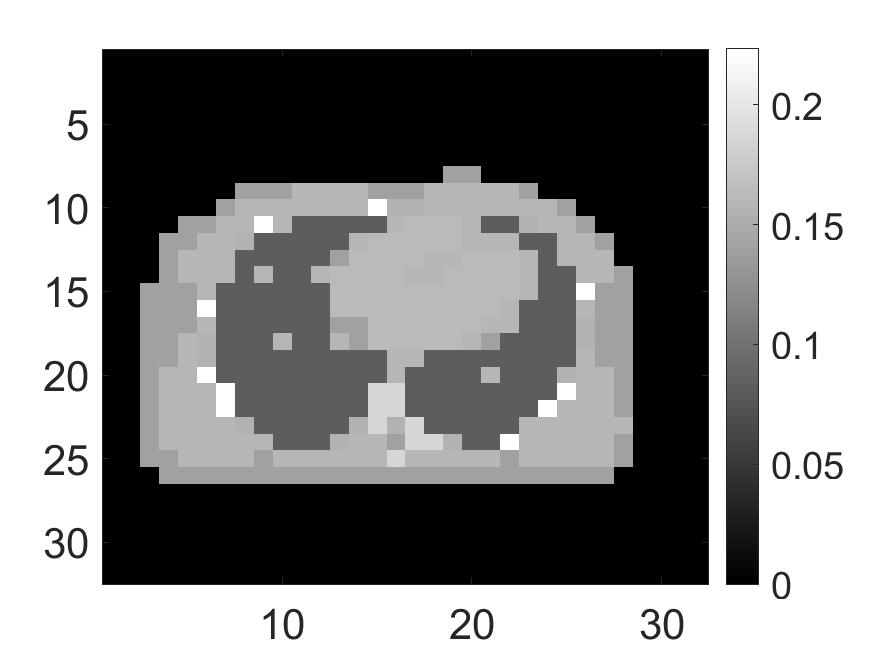}
\label{fig:xcat_attn}
}
\subfloat[]{
\includegraphics[width = 2 in]{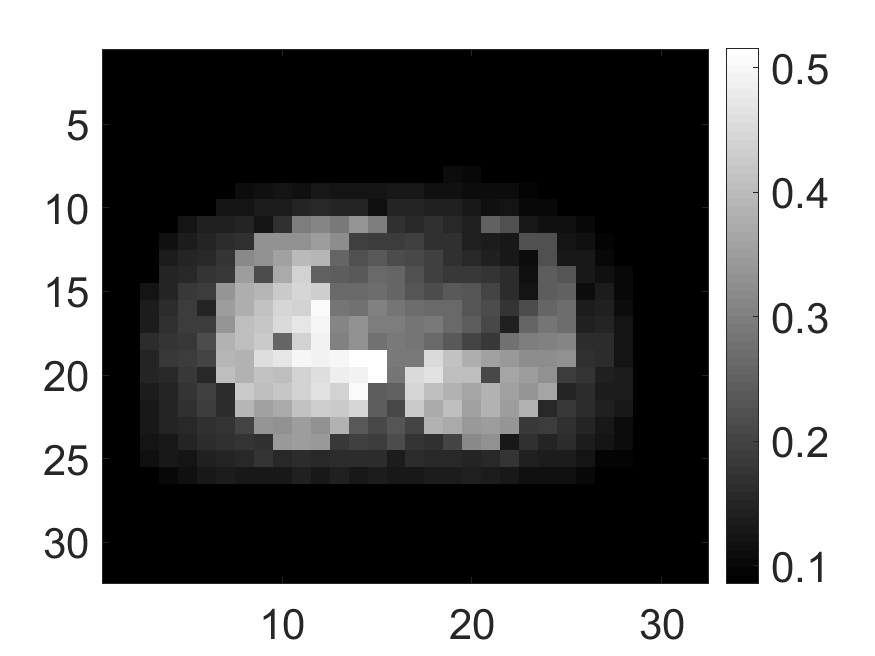}
\label{fig:xcat_std}
}

\label{fig:xcat_act_atn_std}
\hfill
\subfloat[]{
\includegraphics[width = 2 in]{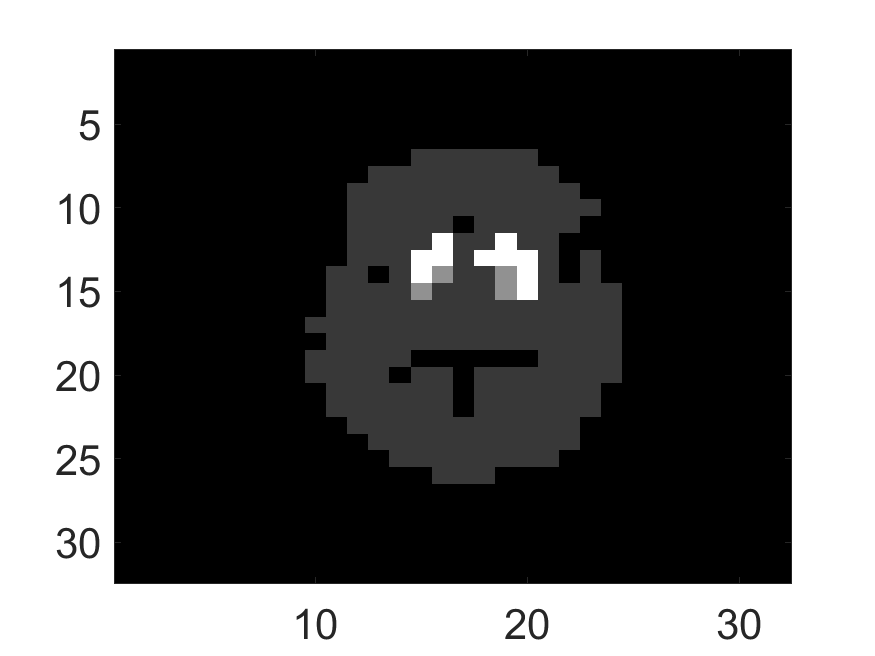}
\label{fig:brain_act}
}
\subfloat[]{
\includegraphics[width = 2 in]{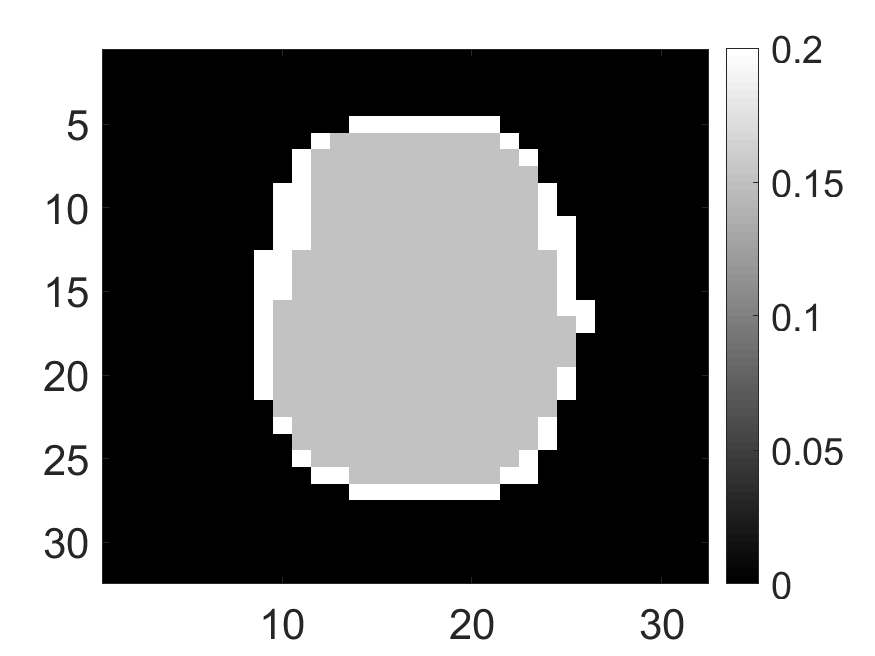}
\label{fig:brain_attn}
}
\subfloat[]{
\includegraphics[width = 2 in]{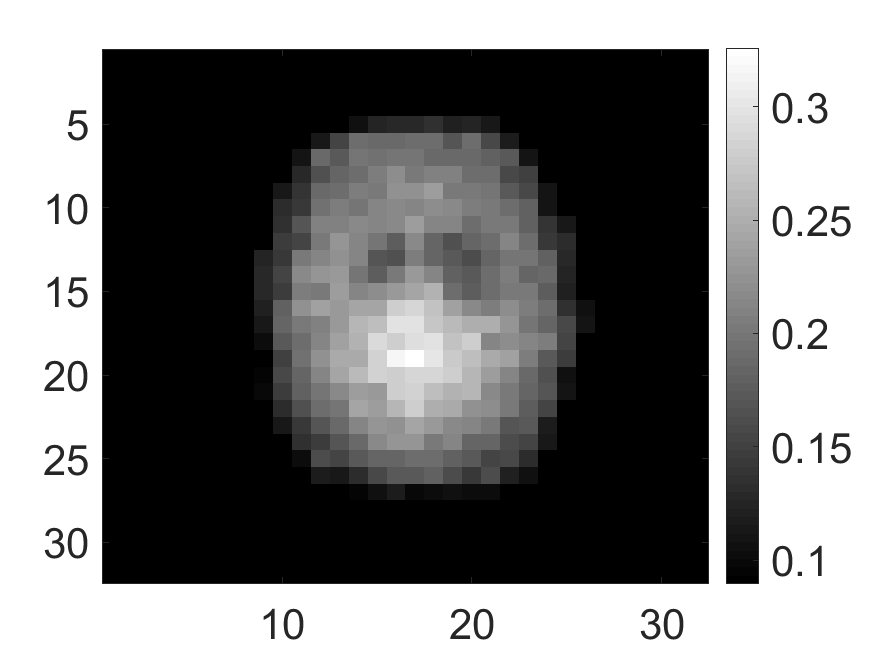}
\label{fig:brain_std}
}

\hfill

\caption{(a) True activity map, (b) true attenuation map and (c) normalized standard deviation of attenuation map for cardiac phantom. (d) True activity map, (e) true attenuation map and (f) standard deviation of attenuation map for brain phantom}
\label{fig:brain_act_atn_std}
\end{figure}

The standard deviation values of activity and attenuation coefficients, using the computed CRB, obtained for each pixel were normalized by dividing by the true value of the activity and attenuation coefficient of that pixel, respectively. This was done for easier interpretation of the results, and in particular to assess whether the standard deviation is smaller than the true value. 
This normalization is especially helpful when analyzing results from phantoms with non-uniform attenuation maps.
The computed normalized standard deviation of attenuation coefficients for the different phantoms when $4 \times 10^5$ detected events were considered is shown in an image format in Figs.~\ref{fig:stddev_phantoms} and \ref{fig:brain_act_atn_std}. In all the results, including those with the cardiac (Fig.~\ref{fig:xcat_std}) and brain (Fig.~\ref{fig:brain_std}) phantoms, the standard deviation of the attenuation coefficient for all the pixels was lower than the true value of the attenuation coefficient. Further, the standard deviation of the attenuation coefficient was low in the pixels that had non-zero activity values.
This observation is most easily apparent in the synthetic phantoms, in particular the single-pixel and multi-pixel phantoms. 
For example, in the single-pixel phantom, the standard deviation was the lowest in the center pixel, the only pixel with activity (Figs.~\ref{fig:stddev_single_vox} and \ref{fig:stddev_single_vox_nu}).
Further, the standard deviation increased radially as we moved away from this pixel for uniform attenuation map (Fig.~\ref{fig:stddev_single_vox}). 
Further, pixels with uptake had a standard deviation much lower than the true value of the attenuation coefficient. 

\begin{figure}[t]
\centering
\subfloat[]{
\includegraphics[width = 2 in]{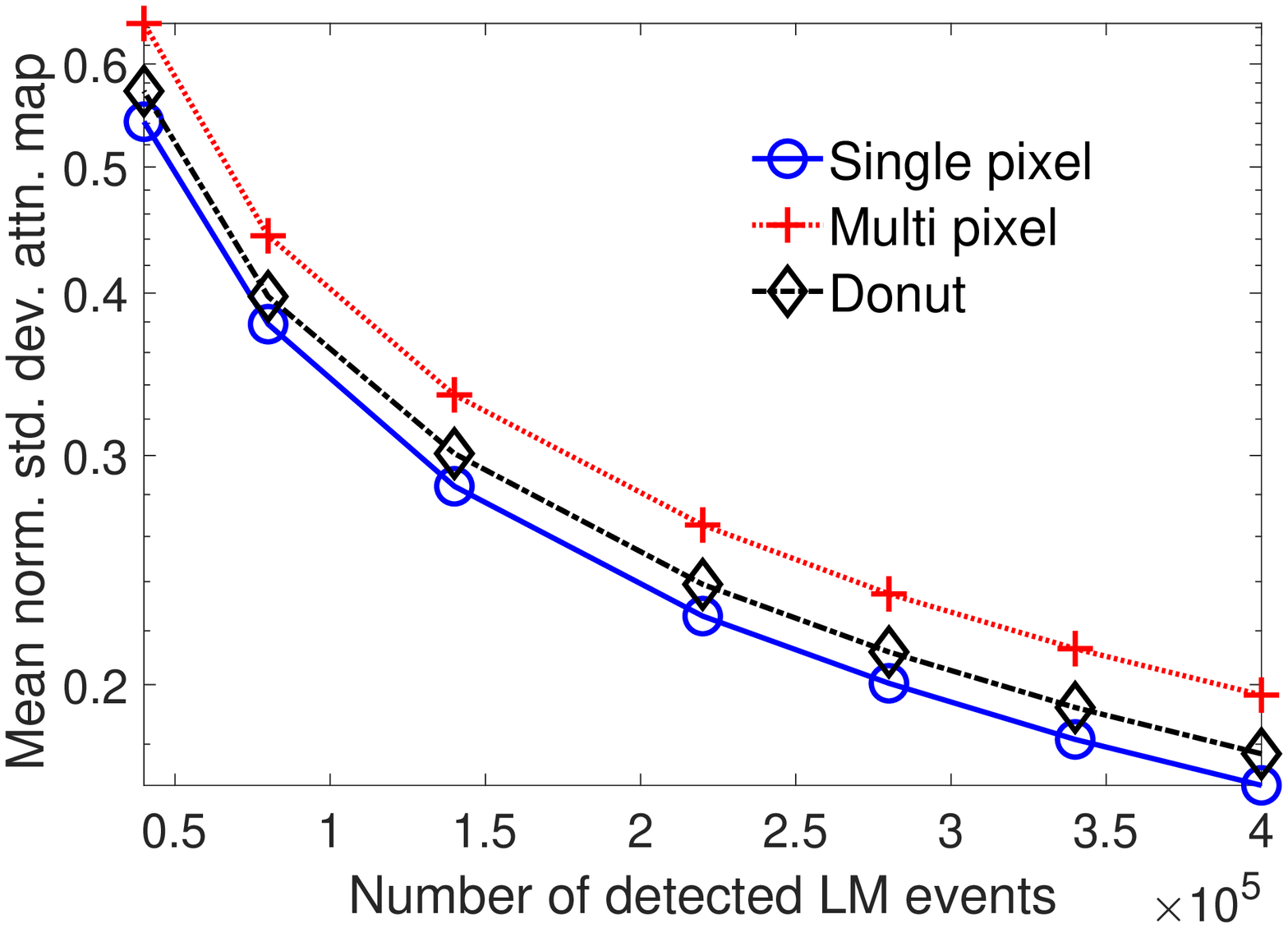}
\label{fig:stddev_attn_allphantoms_uni}
}
\subfloat[]{
\includegraphics[width = 2 in]{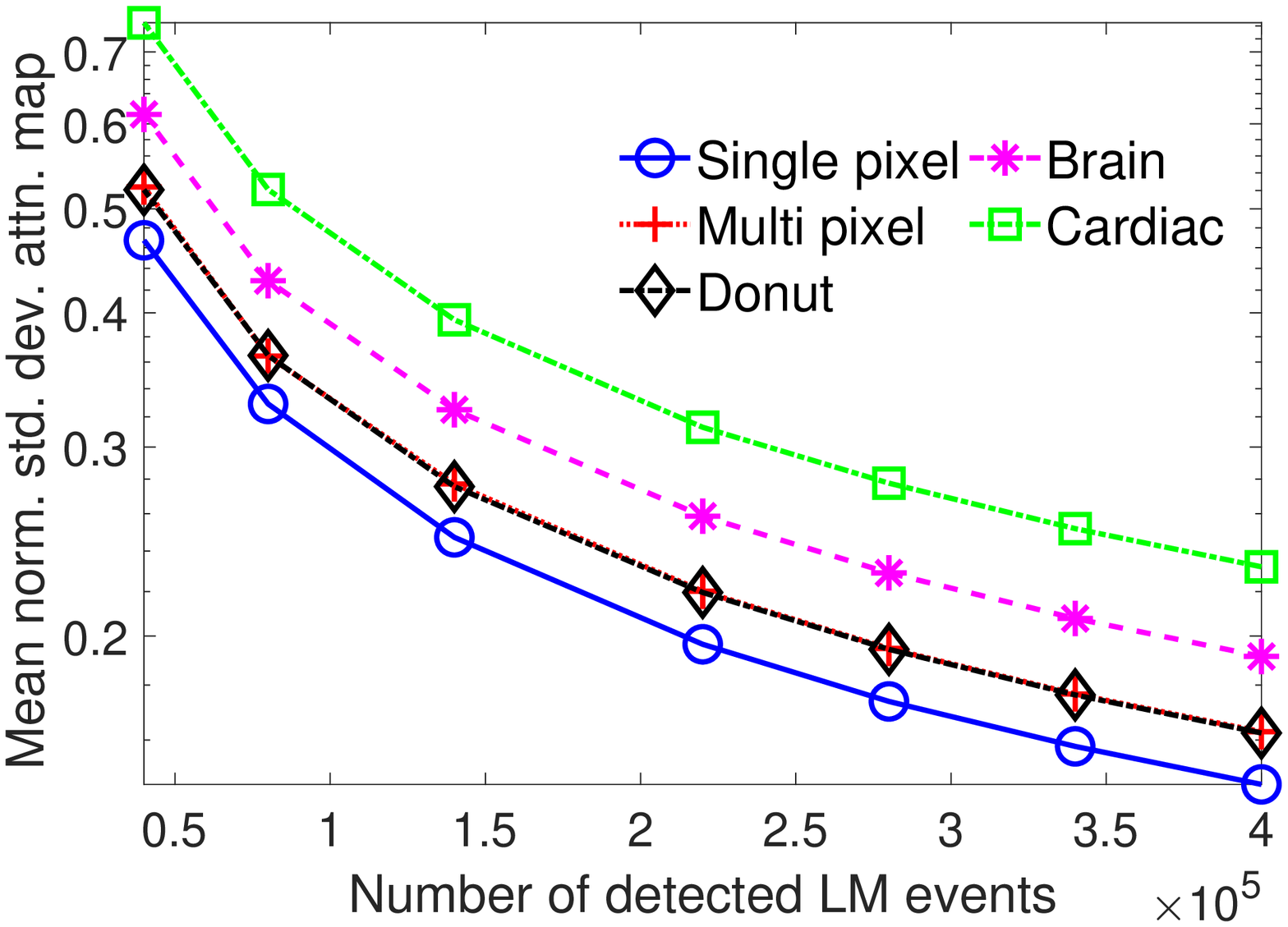}
\label{fig:stddev_attn_allphantoms_nu}
}
\subfloat[]{
\includegraphics[width = 2 in]{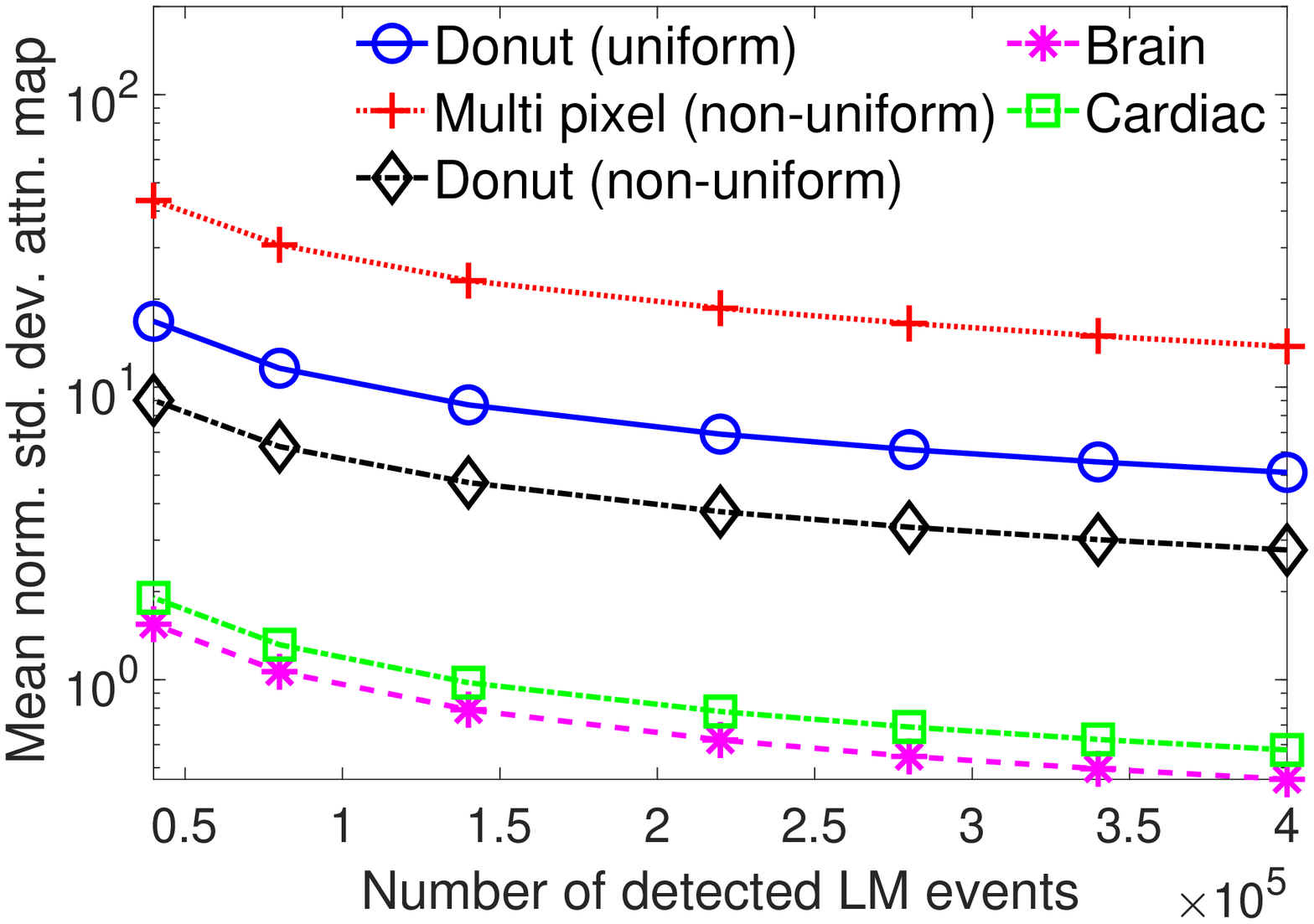}
\label{fig:stddev_attn_allphantoms_nu_anth}
}
\caption{The mean of normalized standard deviation of the attenuation coefficient as a function of the number of detected LM events for (a) synthetic phantoms in uniform attenuation map (b) anthropomorphic phantoms and synthetic phantoms in non-uniform attenuation map and (c) synthetic and anthropomorphic phantoms when only photo-peak events are considered}
\label{fig:stddev_attn_allphantoms}
\end{figure}

\begin{figure}
\centering
\subfloat[]{
\includegraphics[width = 2 in]{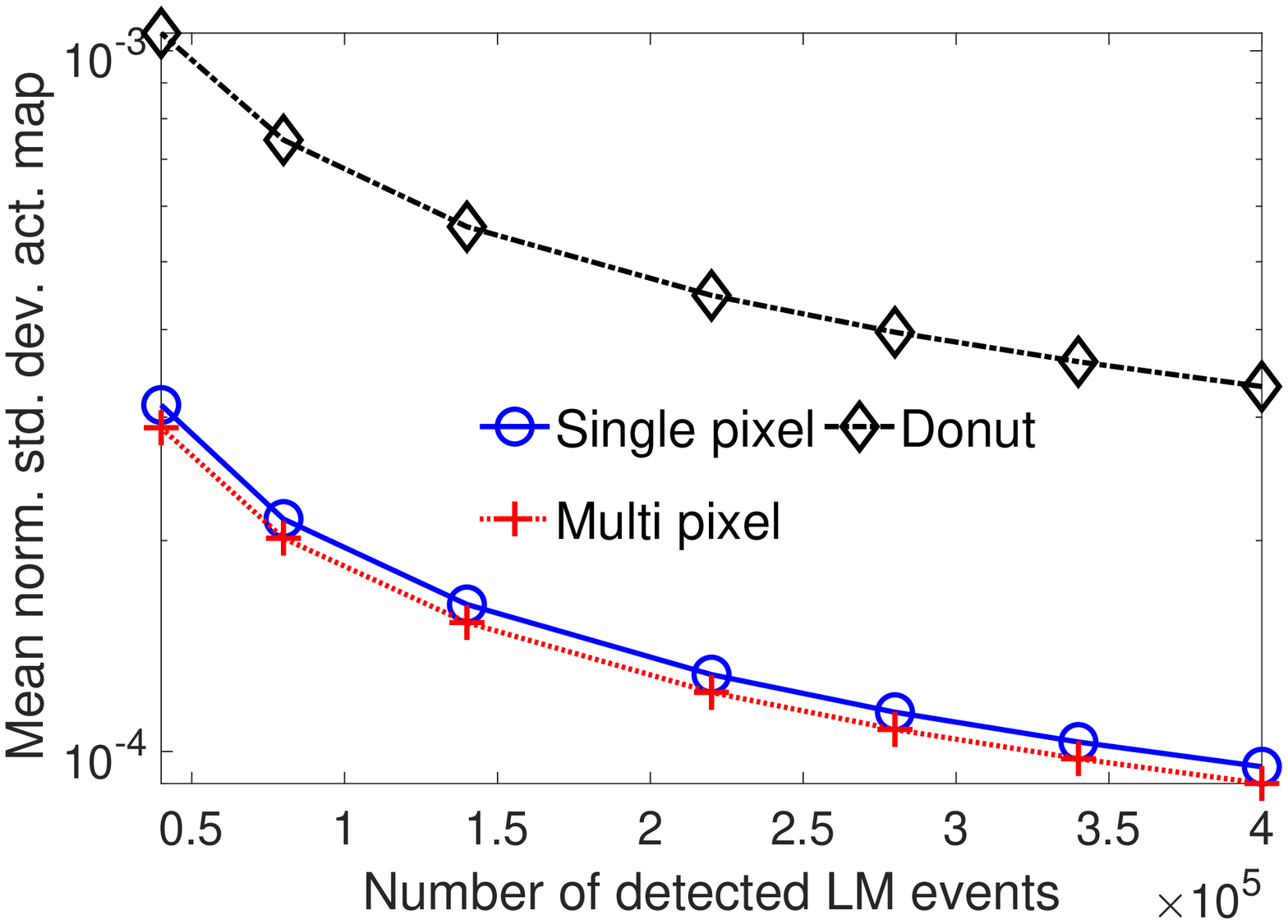}
\label{fig:stddev_act_allphantoms_uni}
}
\subfloat[]{
\includegraphics[width = 2 in]{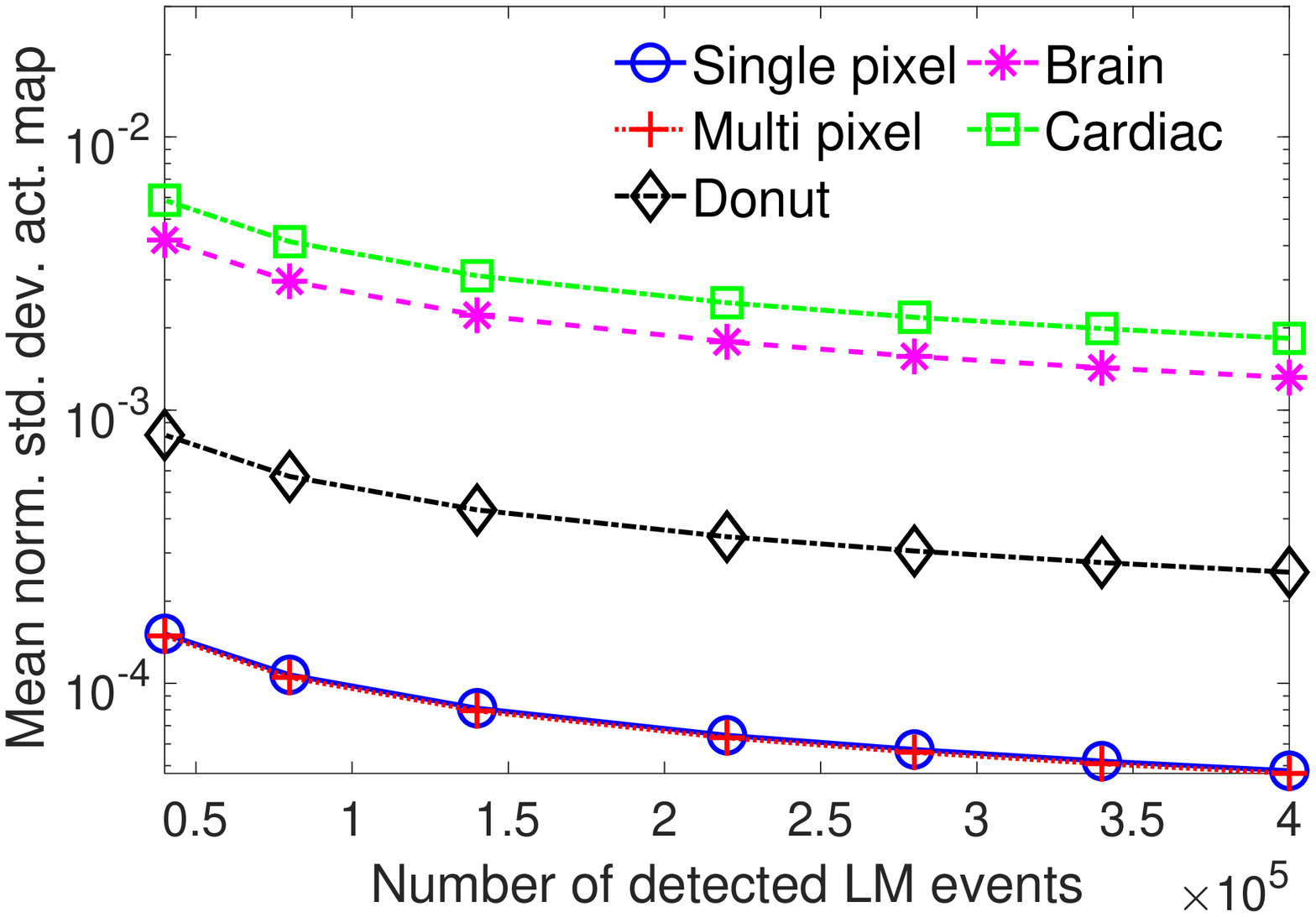}
\label{fig:stddev_act_allphantoms_nu}
}
\subfloat[]{
\includegraphics[width = 2 in]{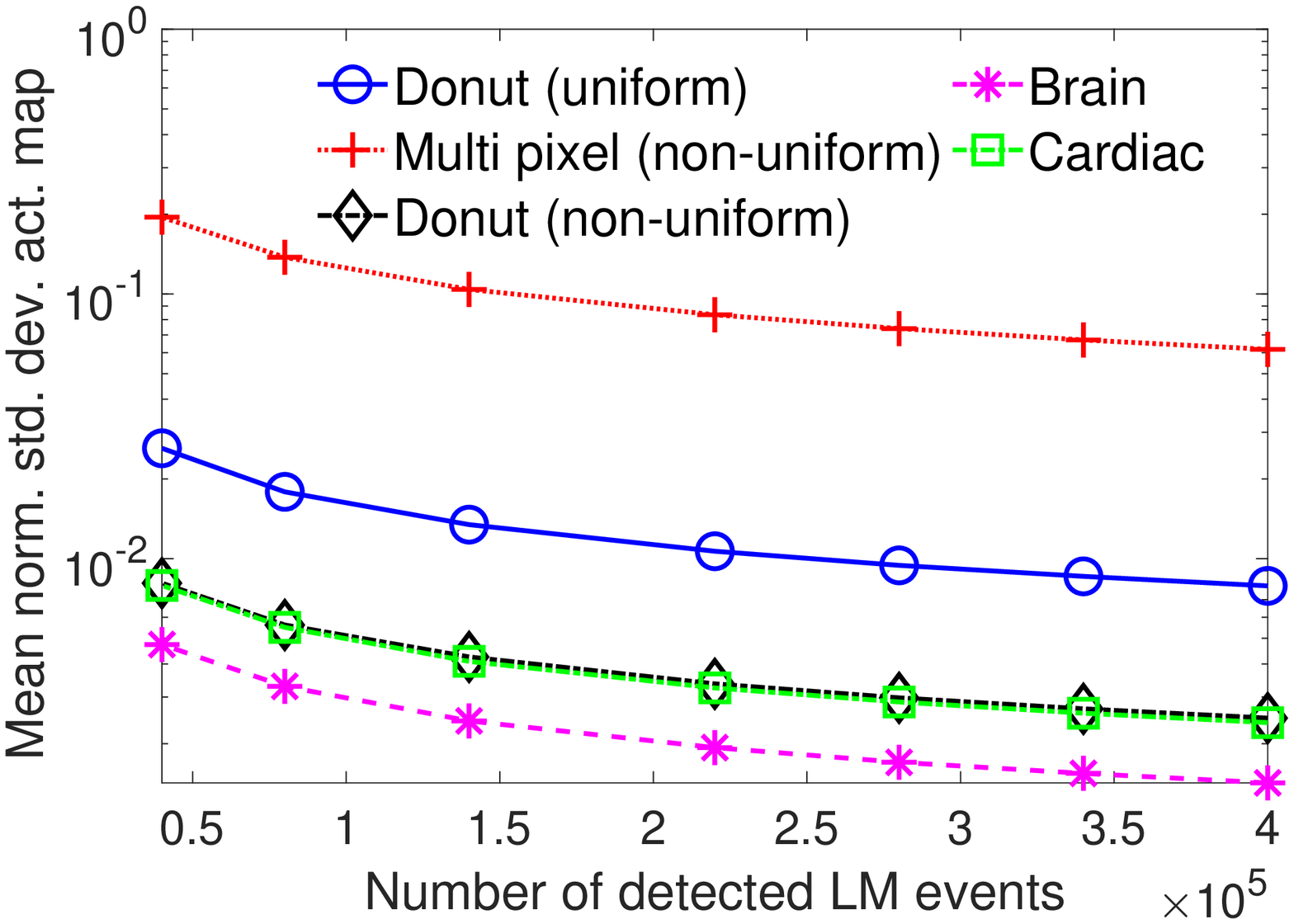}
\label{fig:stddev_act_allphantoms_nu_anth}
}
\caption{The mean of normalized standard deviation of the activity as a function of the number of detected LM events for (a) synthetic phantoms in uniform attenuation map (b) anthropomorphic phantoms and synthetic phantoms in non-uniform attenuation map and (c) synthetic and anthropomorphic phantoms when only photo-peak events are considered}
\label{fig:stddev_act_allphantoms}
\end{figure}

The effect of varying the number of detected LM events on the CRB for the attenuation and activity coefficients is shown in Figs.~\ref{fig:stddev_attn_allphantoms} and \ref{fig:stddev_act_allphantoms}, respectively.
In these plots, we show the mean of the normalized standard deviation of activity and attenuation coefficients over all the pixels having non-zero activity and attenuation, respectively.
%The normalization is again conducted by dividing by the corresponding true value in each pixel.
We observe that the standard deviation values are lower compared to the true values in the mean sense for synthetic and anthropomorphic phantoms at all count levels (Figs.~
\ref{fig:stddev_attn_allphantoms_uni},
\ref{fig:stddev_attn_allphantoms_nu},
\ref{fig:stddev_act_allphantoms_uni} and \ref{fig:stddev_act_allphantoms_nu}). 
In contrast, when using only the data in PP window (Figs.~\ref{fig:stddev_attn_allphantoms_nu_anth} and \ref{fig:stddev_act_allphantoms_nu_anth}), the standard deviation for the activity and attenuation coefficients were infinity for the single-pixel and the multi-pixel phantoms in uniform
attenuation, irrespective of the amount of activity.
Thus, inclusion of scattered photons had a significant impact on the CRB for these phantoms. 
For the other phantoms too, while the PP data provided finite CRB, but that was much higher than the true value of the attenuation coefficient (Fig.~\ref{fig:stddev_attn_allphantoms_nu_anth}). When the scattered photons were included, the CRB was significantly lowered, and became smaller than the true attenuation
coefficient (Fig.~\ref{fig:stddev_attn_allphantoms_nu}). 
All these results demonstrate that the scattered photons contain information to estimate the attenuation coefficient. 

We observe in Figs.~\ref{fig:stddev_attn_allphantoms_uni},~\ref{fig:stddev_attn_allphantoms_nu},~\ref{fig:stddev_act_allphantoms_uni} and \ref{fig:stddev_act_allphantoms_nu} that as the number of detected counts increases, the standard deviation reduces for all the phantoms. 
An increase in the number of detected photons also corresponds to an increase in the number of scattered photons. This provides further evidence that scattered photons contain information about the attenuation coefficient.

%When using only the photopeak data, the standard deviation for the attenuation coefficients was six orders of magnitude higher than the calculated standard deviation when scattered data was included.

\begin{figure}
\centering
\subfloat[]{
\includegraphics[width = 2 in]{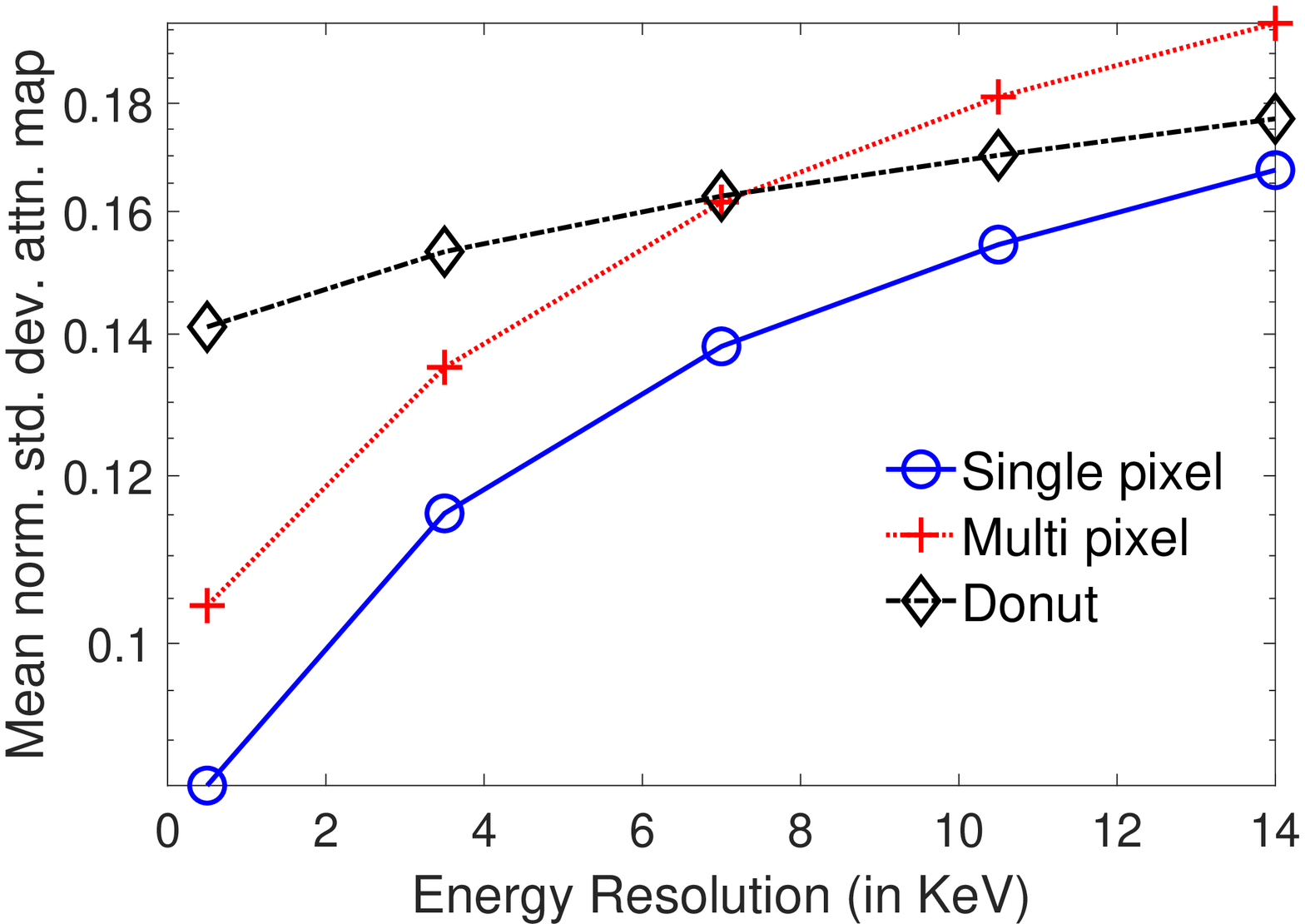}
\label{fig:stddev_attn_vary_energyres_uni}
}
\subfloat[]{
\includegraphics[width = 2 in]{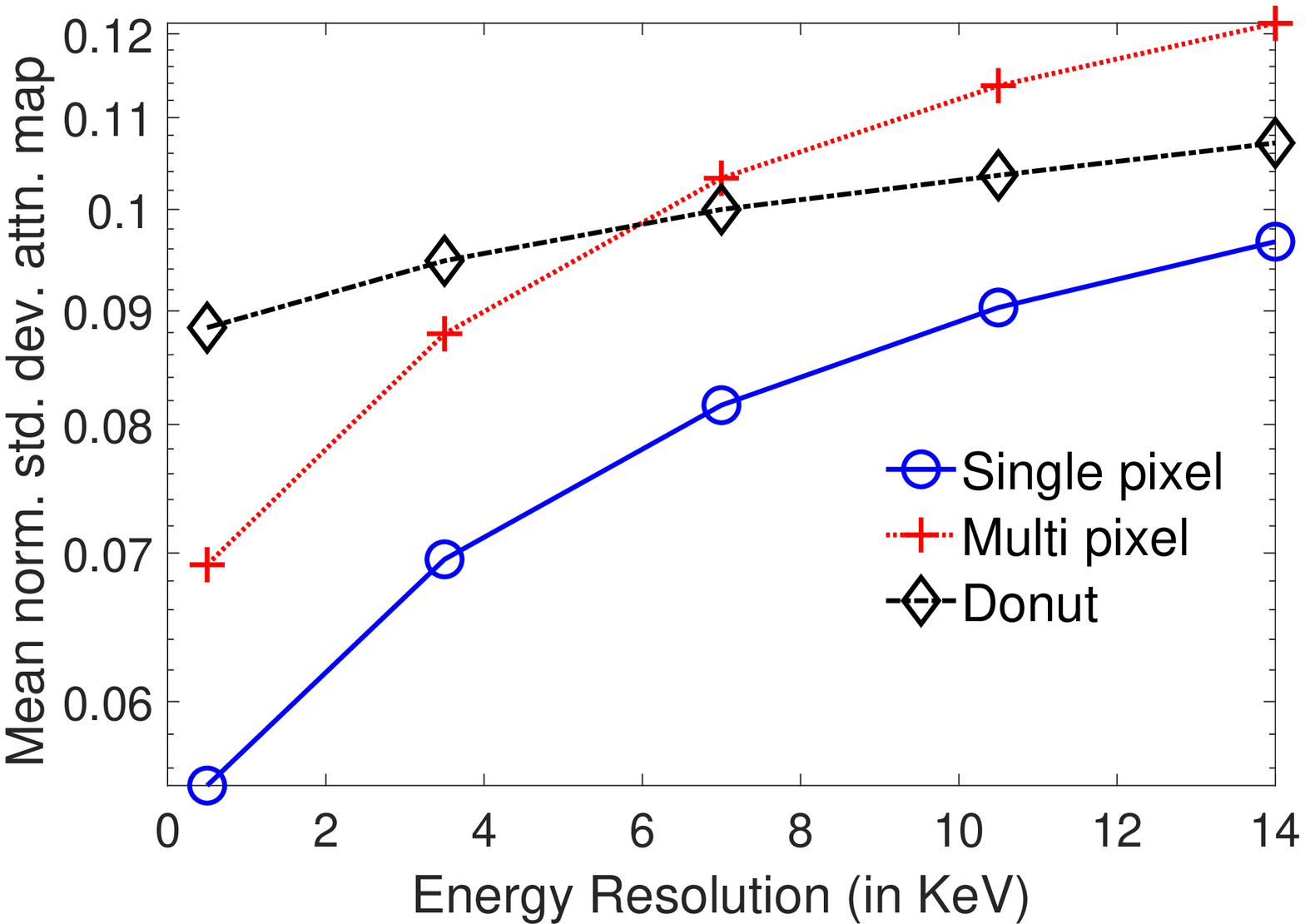}
% For reference, the standard deviation obtained using only the photopeak data with 0.5~keV energy resolution is also plotted.
\label{fig:stddev_attn_vary_energyres_nu}
}
\subfloat[]{
\includegraphics[width = 2 in]{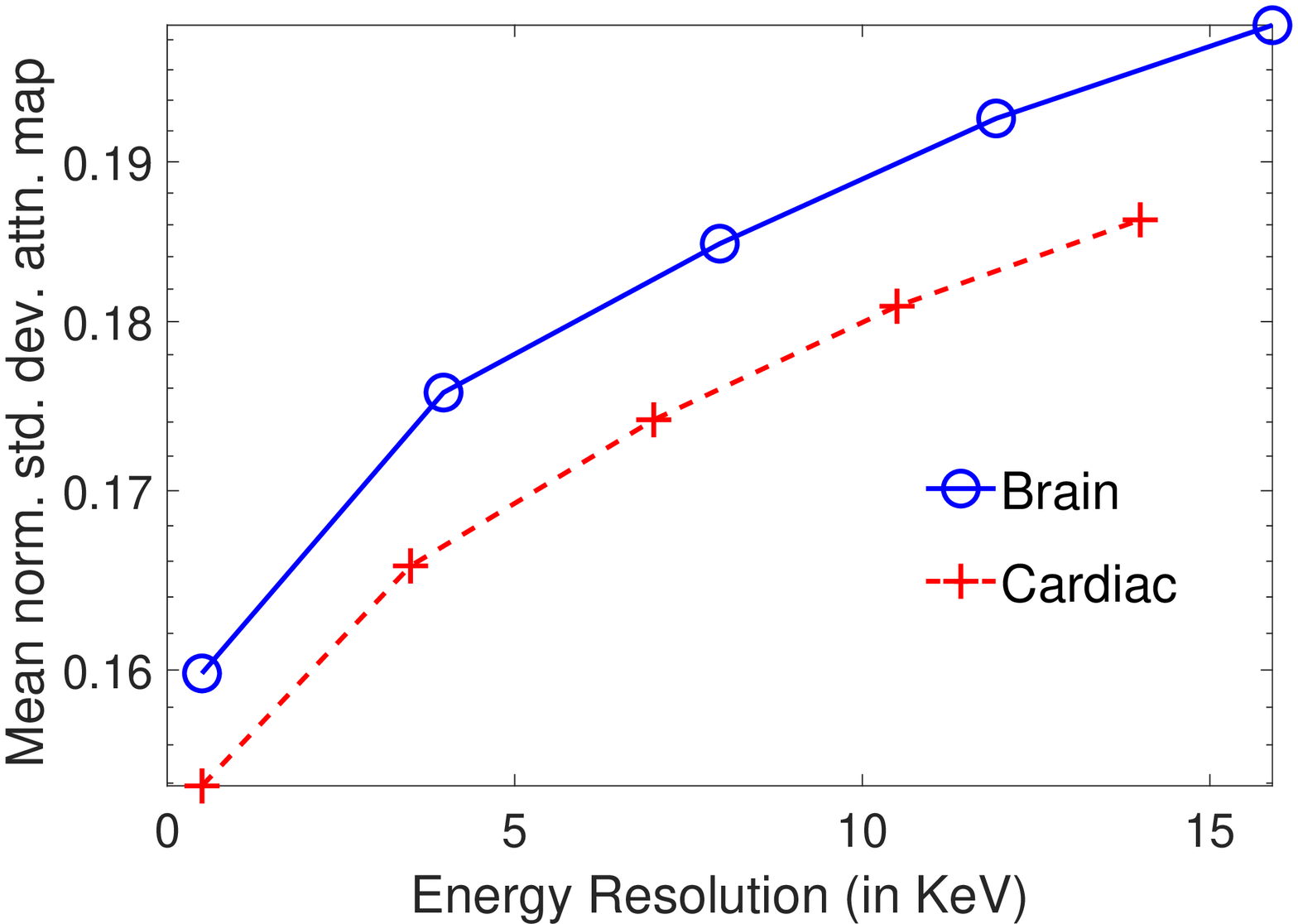}
% For reference, the standard deviation obtained using only the photopeak data with 0.5~keV energy resolution is also plotted.
\label{fig:stddev_attn_vary_energyres_nu_anth}
}
\caption{The mean of normalized standard deviation of the attenuation coefficients as a function of the energy resolutions for different synthetic phantoms with (a) uniform and (b) non-uniform attenuation map, and (c) anthropomorphic phantoms. The normalized standard deviation was calculated for $4\times10^5$ detected LM events}
\label{fig:stddev_attn_vary_energyres}
\end{figure}

In Fig.~\ref{fig:stddev_attn_vary_energyres}, the mean of normalized standard deviation of the attenuation distribution are plotted as a function of  different energy resolutions of the SPECT system.  
We observed that as the energy resolution improved, the standard deviation of the attenuation coefficients reduced. This finding shows that the energy information of the photons contains information that helps with estimating the attenuation distribution. Further implications of this result are in Discussions section. 

\begin{figure}[h!]
	\centering
	\subfloat[]{
		\includegraphics[width=.42\textwidth]{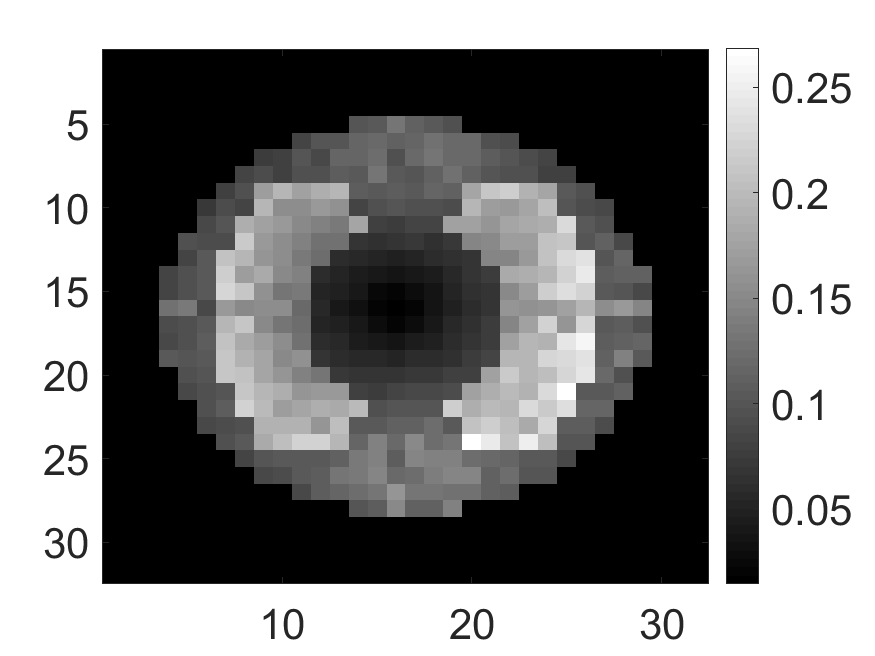}
		\label{fig:sv_ds}
	}
	\subfloat[]{
		\includegraphics[width=.42\textwidth]{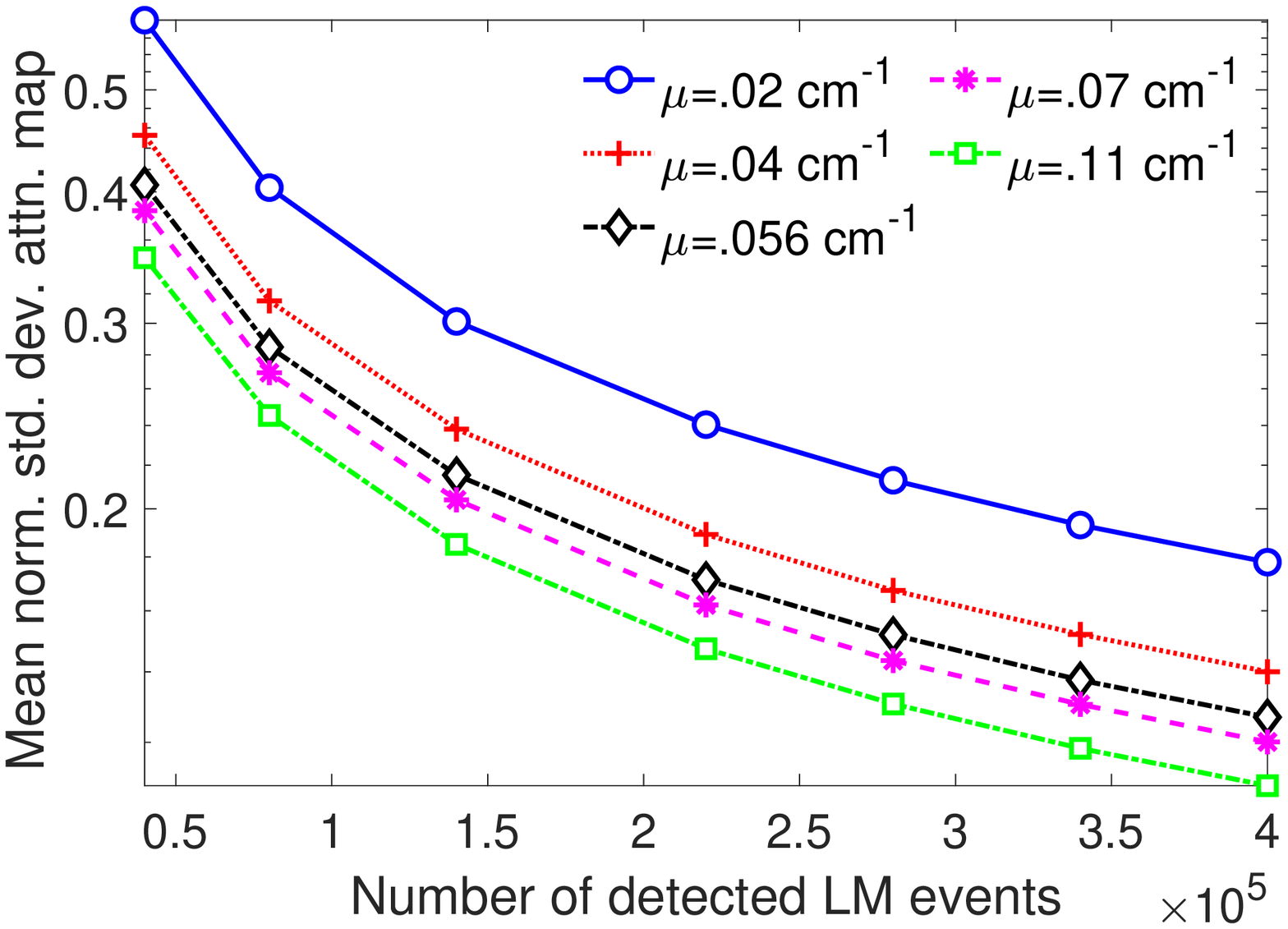}
		\label{fig:s2_s1_dm_nr}
	}	
	
	\caption{(a) Normalized standard-deviation map of attenuation coefficients for single-pixel phantom in non-uniform medium with $4\times10^5$ counts, as shown in Fig.~\ref{fig:nu_true}, where up to two scatter events are considered and (b) mean of normalized version of the CRB-derived standard deviation as a function of lung attenuation and different photon counts in non-uniform phantom for the dual-scatter case.}
\label{fig:all_s2}
\end{figure}

The normalized standard deviation of attenuation map when up to second-order scattered events are considered is shown in Fig.~\ref{fig:sv_ds} for the single-pixel activity phantom in the non-uniform attenuation medium. We observe that the standard deviation values for all the pixels are lower than the true attenuation coefficient, similar to the single-scatter case. In fact, the standard-deviation map is very similar to case where up to first-order scattered events were
considered (Fig.~\ref{fig:stddev_single_vox_nu}). This indicates that photons that have scattered at most twice also contain information to estimate the attenuation distribution. The value of the mean of normalized standard deviation of the attenuation coefficient as a function of the lung attenuation are shown in Fig.~\ref{fig:s2_s1_dm_nr} for dual-scatter case. The value decreases on increasing the attenuation coefficient for all count levels. These results indicate that even when the proportion of dual-scattered photons increases, the scattered-photon data contains information to estimate the attenuation distribution.

%These results demonstrated that the scattered photons contain information about the attenuation distribution.
%The above three set of results tell us about the standard deviation of the attenuation map coefficients. 
%From the inverse of the FIM, we can also obtain the standard deviation of the activity map. 
%We computed this standard deviation over all the voxels, determined the mean of this standard deviation map, and plotted this as a function of the activity. 
%The results are as shown in Fig.~\ref{stddev_act_vary_act}. 
%We again observe that as the activity increases, or equivalently, the number of detected photons increases, the standard deviation on the estimate of the activity reduces, which is an expected result. 
%Also, the single-voxel phantom has a lower standard deviation as compared to the uniform phantom. 
%This is also expected, since the relative number of photons giving information about the activity in a particular voxel is higher for the single-voxel phantom. 

%\begin{figure}
%\centering
%\includegraphics[width = 4 in]{stddev_act_vary_act.eps}
%\caption{The mean standard deviation of the activity coefficients averaged over the phantom as a function of the total activity in the phantom, for both uniform and single-voxel phantoms.}
%\label{stddev_act_vary_act}
%\end{figure}

\section{Discussions}
\label{sec:Discussions}
Overall, the presented results show that photons that have scattered at most twice and are acquired in LM format contain information to jointly estimate the activity and attenuation coefficient. In particular, results in Figs.~\ref{fig:stddev_phantoms} and \ref{fig:brain_act_atn_std} show that for synthetic as well as anthropomorphic digital phantoms, the standard deviation of the attenuation coefficient is lower than the true value.
Additionally, while considering up to first-order scatter, CRB is usually lower in the pixels having non-zero activity uptakes and increasing the number of detected counts improves the information content. Similar results were obtained in the case where second-order scattered photons were considered for a single-pixel phantom.

\begin{figure}
\centering
\subfloat[]{
\includegraphics[width = 2.4 in]{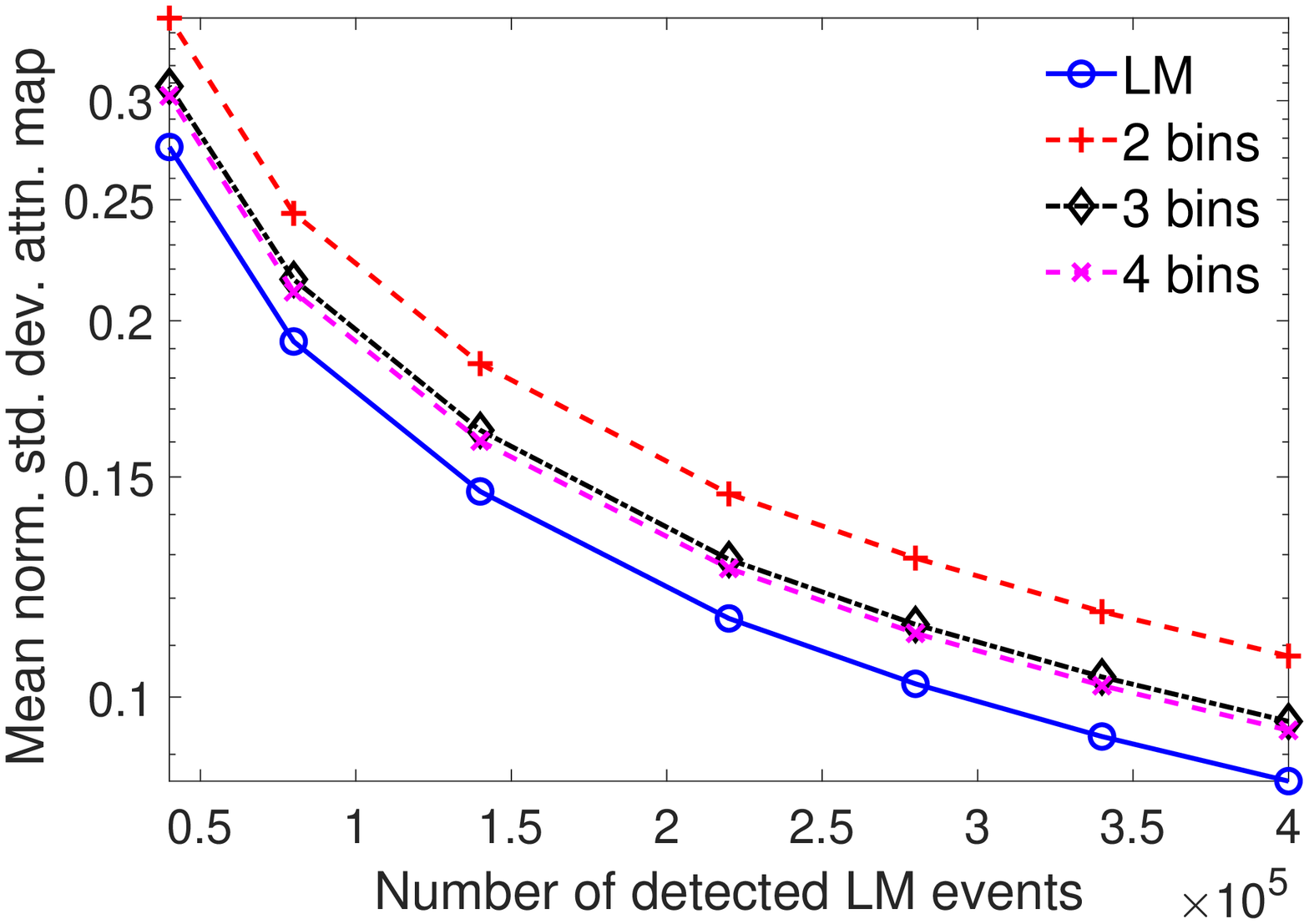}
\label{fig:stddev_attn_vary_energybins_sv}
}
\subfloat[]{
\includegraphics[width = 2.4 in]{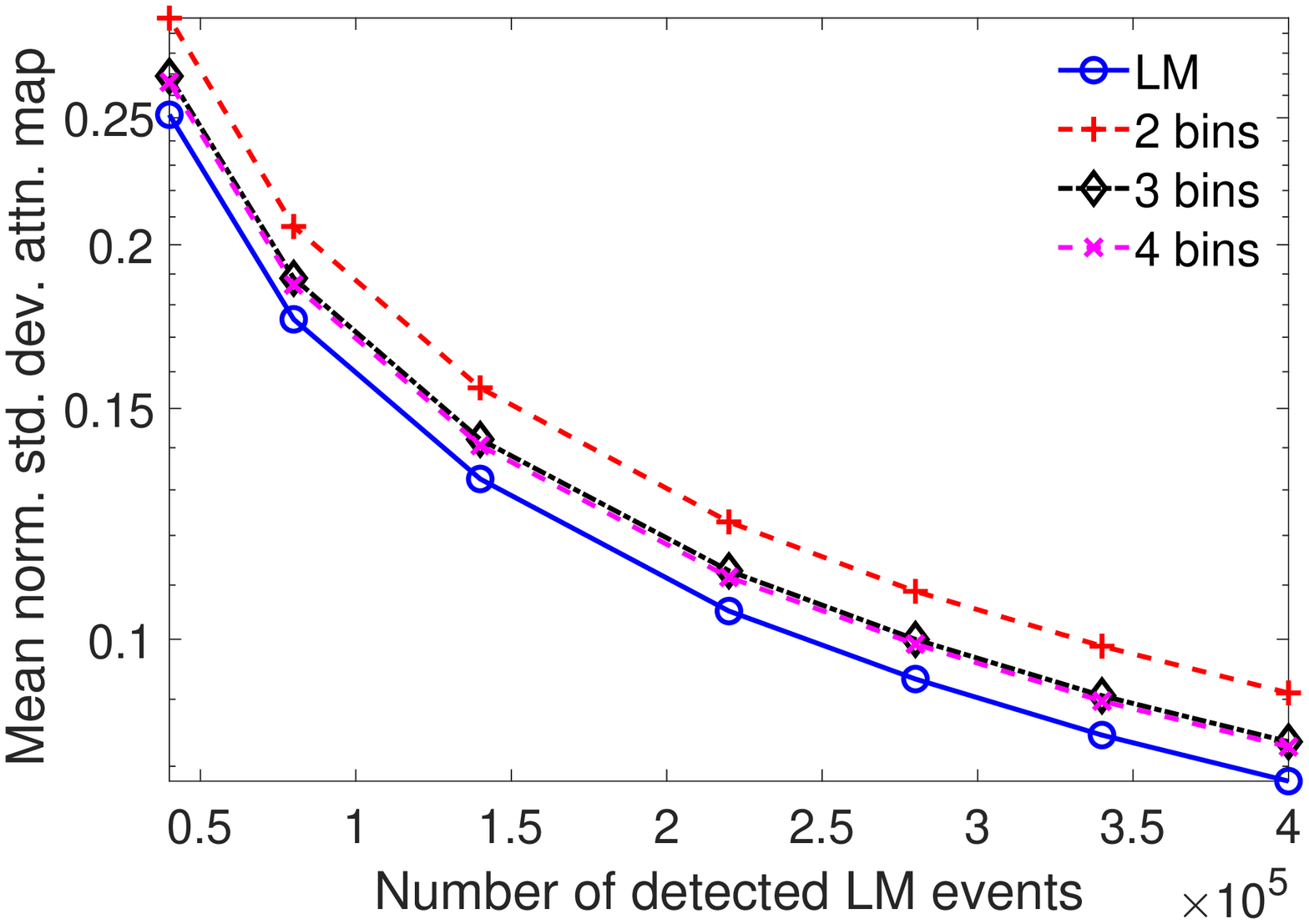}% For reference, the standard deviation obtained using only the photopeak data with 0.5~keV energy resolution is also plotted.
\label{fig:stddev_act_vary_energybins_sv}
}
\caption{The mean of normalized standard deviation of the attenuation coefficient as a function of the number of LM events for single-pixel phantom with (a) uniform and (b) non-uniform attenuation map. The results are shown for LM emission data and for cases where the emission data was binned into different number of energy bins.}
\label{fig:stddev_attn_vary_act_bins}
\end{figure}

The results in Fig.~\ref{fig:stddev_attn_vary_energyres} showed the importance of energy attribute in estimating the attenuation coefficient. We also conducted a study to assess the impact of binning the energy attribute on the CRB of the attenuation coefficients. To study this effect explicitly, the scintillation detector was assigned a very high energy resolution of 0.5 keV. 
% storing the energy attribute in LM format resulted in more information for the joint activity-attenuation estimation task in comparison to if the energy attribute was binned. 
In the binning process, the entire LM data was binned into different number of bins based on the energy value. 
For each configuration, we set the range of energy bin values such that approximately similar number of photons were present in each bin.
All photons with energies within a bin were assigned the same energy as at the center of the bin.
The CRB for the binned and LM data were compared.
The mean of normalized standard deviation of the attenuation coefficient for single-pixel phantom with LM data and data where the energy attribute was binned into different number of bins is plotted in Fig.~\ref{fig:stddev_attn_vary_act_bins}.
%To explicitly study the information loss due to the process of binning the energy attribute, we considered a scintillation detector with a very high energy resolution of $0.5~\mathrm{keV}$.
We observed that as the number of energy bins increased, the standard deviation of the attenuation coefficient reduced. 
Of most importance, LM data yielded a lower standard deviation for the attenuation coefficient in comparison to even when up to four energy bins were considered.
These results are in agreement with other recently conducted studies \cite{Jha:15:pmb,Jha:15:spie,clarkson2019quantifying,henscheid2017evaluation,ding2017null}, which have all shown that binning of LM attributes leads to loss of information. 
%For example, assume a SPECT system consisting of $M_1$ detector pixels and acquiring data over $M_2$ projection angles, imaging an object discretized into $N$ voxels.
%The forward operator for this system would have a size $M_1 \times M_2 \times N$

The results in Fig.~\ref{fig:stddev_attn_vary_energyres} suggest that detectors having better energy resolution can improve the joint estimation of the activity and attenuation distribution. 
In this context, the emerging solid-state detectors such as the Cadmium-Zinc-Telluride (CZT) detectors currently provide an energy resolution of 6\%, and could theoretically provide up to 1.5\% energy resolution \cite{GE_Whitepaper_CZT}.
Our results show that such an increase in energy resolution could improve the joint activity-attenuation estimation capability of these systems.
This is highly significant since several CZT-based systems for cardiac imaging, such as those from Spectrum Dynamics (DSPECT) and GE (NM 530c) have high sensitivity, high energy, temporal, and spatial resolution.
Further, these systems have demonstrated capability to obtain low-dose SPECT images. 
Some of these solid-state systems are also lightweight and portable, such as the Cardius XPO-M system from Digirad, enabling mobile SPECT imaging in remote locations. 
However, these systems often do not have CT imaging capability. 
A recent study has shown that ASC leads to improved diagnostic accuracy with these solid-state-detector systems \cite{Caobelli:16}. 
Thus, enabling ASC using only SPECT emission data for these systems would have significant impact.

A limitation of our study is that we consider photons that scatter at most twice. While the theoretical formalism that we have developed provides the mathematical expressions to conduct such a study, we were limited by the computational requirements. 
%Our results show that single and dual-scatter photons provide information about the attenuation coefficients. Further, the result in Fig. \ref{fig:s2_s1_dm_nr} does show that as the attenuation coefficient increased, the amount of information present in the dual scattered-photons also continued to increase. However, as the attenuation coefficient increases, photons that scatter more than twice also increases, and our analysis does not consider these photons. 
The percent of photons that scatter more than twice in clinical SPECT imaging is relatively small \cite{kojima1993effect,de2001rapid}, but nevertheless, evaluating the information content of these photons is an important research frontier, with applications also in other imaging modalities where incoherent scatter occurs. 

Another limitation of our study is that while we studied the performance of our method with different kind of phantoms, including anthropomorphic phantoms, in all the cases, the phantoms were discretized over a $32 \times 32$ grid. However, in SPECT, the reconstructed images are discretized over a $64 \times 64$ or $128 \times 128$ grid.
This limitation arises due to the computational and memory requirements of the software. 
Advances in computational hardware provide a mechanism to address this challenge. 
At the same time, our results do indicate the possibility to reconstruct the attenuation distribution over a low-resolution $32 \times 32$ grid. Thus, one possibility is that this low-resolution attenuation map can be interpolated to a higher-resolution attenuation distribution, which could then be used for attenuation compensation. Evaluating the efficacy of such an approach would require studies that objectively assess the quality of the reconstructed activity images on the clinical task. 
A third limitation is that the study was conducted for a 2-D SPECT system with 2-D phantoms. 
%Thus, the applicability of these results is when we have a 2D reconstruction method.
However, the theoretical treatment is completely general and implementing this study for a 3-D SPECT system is an important direction of future research.
Finally, in our method, a few processes such as inter-septal and inter-crystal scatter are not modeled. Modeling these additional processes will make the study even more realistic.

Results from this study motivate application of this method to
anthropomorphic physical-phantom studies and to patient data. These
studies will provide further insights on the information content of SPECT emission
data for joint reconstruction in clinically more realistic settings. 
%Promising results will motivate designing methods for joint reconstruction of activity and attenuation distribution using only the SPECT emission data.
The results from this study also motivate the development of methods to jointly estimate activity and attenuation distribution using only the SPECT emission data, especially if the computational requirements of processing LM data can be reduced. Efforts in the direction of reducing these computational requirements are currently underway \cite{caucci2019towards,asheq_2}.
%Motivated by this fact, we tried to observe the effect of number of list-mode events that are scattered multiple times while confining our information content analysis for single-scatter. For this purpose, we varied the attenuation coefficient of the lung-like region shown in Fig.~\ref{fig:nu_true} from $0.02~\cmeter^{-1}$ to $0.11~\cmeter^{-1}$. The mean standard deviation of the attenuation coefficient is plotted in Fig.~\ref{fig:msmc_nu_difat} as a function of different attenuation coefficient in the lung-like region for different phantoms. The standard deviation of attenuation coefficient converges as number of multi-scattered event increases.     

%The presented results motivate application of this method to computational and physical-phantom studies with anthropomorphic phantoms.
%These studies will provide further insights on the information content of SPECT emission data for joint reconstruction in clinically more realistic settings.  
%Promising results will motivate and provide strong justifications for designing and validating methods for joint reconstruction of activity and attenuation distribution using only the SPECT emission data.  

\section{Conclusions}
\label{sec:Conclusions}
We have investigated the information content of LM SPECT emission data, which includes the scattered-photon data and the energy attribute for each detected photon, for the task of jointly estimating the activity and attenuation distributions.
For this purpose, we developed a Fisher-information-based method that yielded the CRB for the activity and attenuation coefficients from SPECT LM data for the joint estimation task. 
In the process, we also proposed a path-based formalism to process the LM scattered-photon data.
Computational experiments with a simulated 2-D SPECT imaging system, and synthetic and anthropomorphic digital phantoms, for different photon count levels, demonstrated that photons that had scattered at most once contain information about the attenuation coefficients.
Similar results were observed for the case when photons that scattered at most twice were considered.
The standard deviation of the attenuation coefficient was lower than the true attenuation coefficient value for clinical count levels.
% providing evidence that the entire LM SPECT emission data could be used to estimate the attenuation distribution.
Further, improving the energy resolution of the SPECT system resulted in more information about the attenuation coefficients.
%Additionally, the energy attribute stored in LM format provided more information to estimate the attenuation coefficient in comparison to when stored in binned format.
Overall, the results provide promising evidence that the LM SPECT emission data, including the scattered-photon data that includes the energy attribute, contain information to jointly estimate the activity and attenuation distributions.

\section*{Acknowledgment}		
This work was supported by National Institute of Biomedical Imaging and Bioengineering of National Institute of Health under grant number R21-EB024647, R01-EB016231, R01-EB000803 and SNMMI Bradley-Alavi Fellowship. 
Support is also acknowledged from the NVIDIA GPU grant. 
The authors thank Drs.~Harrison H. Barrett, Brian Hutton, and Jonathan Links for helpful discussions. 
\section*{Appendix A: Deriving elements of FIM}
To derive the elements of the FIM, we start from Eq.~\eqref{diff_likelihood_lambda} and Eq.~\eqref{diff_likelihood_mu}.
Differentiating Eq.~\eqref{diff_likelihood_lambda} with respect to the activity in the $q'^{\th}$ voxel, $\lambda_{q'}$ yields
\begin{align}
& \frac{\partial^2 \L}{\partial \lambda_{q'} \partial \lambda_q } = - \sum_{j=1}^J \frac{\{\sum_{\P_q} \pr(\bAhat_j | \P) s_{\eff} (\P) \} \{\sum_{\P_{q'}} \pr(\bAhat_j | \P) s_{\eff} (\P) \}}{\{\sum_{\P} \pr(\bAhat_j | \P) \lambda(\path) s_{\eff} (\P) \}^2 }.
\label{double_diff_likelihood_lambda}
\end{align}
Differentiating Eq.~\eqref{diff_likelihood_mu} with respect to $\mu_{q'}$ gives
\begin{align}
\frac{\partial^2 \L}{\partial \mu_{q'} \partial \mu_q } &=  - \sum_{j=1}^J \left[ \frac{ \left\{ \sum_{\P} \pr(\bAhat_j | \P) \lambda(\P)  s_{\eff} (\P)  \left[ -\Delta_q(\path) + \frac{\zeta_q(\path)}{\mu_q} \right] \right\} }{\{{\sum_{\P} \pr(\bAhat_j | \P) \lambda(\P) s_{\eff} (\P)}\}^2} \times \right. \nonumber \\
& \quad \left \{ \sum_{\P} \pr(\bAhat_j | \P) \lambda(\P)  s_{\eff} (\P)  \left[ -\Delta_{q'}(\path) + \frac{\zeta_{q'}(\path)}{\mu_{q'}} \right] \right \} + \nonumber \\
& \quad \left. \frac{\sum_{\P} \pr(\bAhat_j | \P) \lambda(\P) s_{\eff} (\P)  \left[ -\Delta_q(\P) + \frac{\zeta_q(\path)}{\mu_q} \right] \left[ -\Delta_{q'}(\P) + \frac{\zeta_{q'}(\path)}{\mu_{q'}} \right]}{\sum_{\P} \pr(\bAhat_j | \P) \lambda(\path) s_{\eff} (\P)} \right]  - \nonumber \\
& \quad T \sum_{\path} \lambda(\path) s_{\eff}(\path) \left[ -\Delta_q(\P) + \frac{\zeta(\path)}{\mu_q} \right] \left[ -\Delta_{q'}(\P) + \frac{\zeta_{q'}(\path)}{\mu_{q'}} \right].
\label{double_diff_loglikelihood_mu}
\end{align}
Similarly, differentiating Eq.~\eqref{diff_likelihood_lambda} with respect to $\mu_{q'}$ gives
\begin{align}
& \frac{\partial^2 \L}{\partial \mu_{q'} \partial \lambda_q } = -\sum_{j=1}^J \frac{\left\{\sum_{\P_q} \pr(\bAhat_j | \P) s_{\eff} (\P) \right\} \left\{\sum_{\P} \pr(\bAhat_j | \P) \lambda(\P) s_{\eff} (\P) \left[-\Delta_{q'}(\P) + \frac{\zeta_{q'}(\P)}{\mu_{q'}} \right] \right\}}{\left\{\sum_{\P} \pr(\bAhat_j | \P) \lambda(\path) s_{\eff} (\P) \right\}^2 } + \nonumber \\
& \sum_{j=1}^J \frac{ \sum_{\P_q} \pr(\bAhat_j | \P) s_{\eff} (\P) \left[ -\Delta_{q'} + \frac{\zeta_{q'}(\path)}{\mu_{q'}}\right]  } { \sum_{\P} \pr(\bAhat_j | \P) \lambda(\path) s_{\eff} (\P) } - T \sum_{\P_q} s_{\eff}(\P)  \left[ -\Delta_{q'} + \frac{\zeta_{q'}(\path)}{\mu_{q'}}\right].
\label{double_diff_loglikelihood_cross}
\end{align}
Differentiating Eq.~\eqref{diff_likelihood_mu} with respect to $\lambda_{q'}$ yields
\begin{align}
\frac{\partial^2 \L}{\partial \lambda_{q'} \partial \mu_{q} } = \frac{\partial^2 \L}{\partial \mu_{q} \partial \lambda_{q'}}.
\label{double_diff_loglikelihood_cross2}
\end{align}
To obtain the elements of the FIM for a given value of the activity and attenuation coefficient, the quantities obtained in Eqs.~\eqref{double_diff_likelihood_lambda}-\eqref{double_diff_loglikelihood_cross2} must be averaged with respect to the observed LM data at that value of the activity and attenuation coefficient. 
Before averaging, note that 
\begin{align}
\sum_{\path} \pr(\bAhat_j | \P) \lambda(\path) s_{\eff} (\P) &=   \beta \sum_{\path} \pr(\bAhat_j | \P) \Pr(\path| \emis, \att) = \beta \sum_{\path} \pr(\bAhat_j | \emis, \att),
\label{den_term_double_diff}
\end{align}
where in the second and third steps, Eq.~\eqref{eq:pr_path_2} and the mixture-model definition (Eq.~\eqref{mixture_model}) have been used, respectively.

To evaluate the FIM elements with respect to attenuation coefficients, start from Eq.~\eqref{double_diff_loglikelihood_mu}, and consider the second term in this equation. 
Substitute Eq.~\eqref{den_term_double_diff} in the denominator of the second term in Eq.~\eqref{double_diff_loglikelihood_mu}. 
Next, average over the LM attributes $\hat{A}$. 
To perform the averaging operation over $\hat{A}$, note that $\pr(\hat{\A}| J, \emis, \att) = \pr(\bAhat_1, \bAhat_2, \ldots \bAhat_J | \emis, \att) = \pr(\bAhat_1| \emis, \att) \ldots \pr(\bAhat_j| \emis, \att) \ldots \pr(\bAhat_J| \emis, \att)$, since the $J$ LM events are independent of each other. 
%the second term in Eq.~\eqref{double_diff_loglikelihood_mu} reduces to
%\begin{align}
%& \left\langle \sum_{j=1}^J \frac{\sum_{\P} \pr(\hat{A}_j | \P) \lambda(\P) s_{\eff} (\P)  \left[ -\Delta_q(\P) + \frac{\zeta(\path)}{\mu_q} \right] \left[ -\Delta_{q'}(\P) + \frac{\zeta(\path)}{\mu_{q'}} \right]}{\sum_{P} \pr(\hat{A}_j | \P) \lambda(\path) s_{\eff} (\P)} \right\rangle_{(\hat{\A}, J| \emis, \att)}   = \nonumber \\ 
%& \left\langle \sum_{j=1}^J \left\langle \frac{\sum_{P} \pr(\hat{A}_j | \P) \lambda(\P) s_{\eff} (\P)  \left[ -\Delta_q(\P) + \frac{\zeta(\path)}{\mu_q} \right] \left[ -\Delta_{q'}(\P) + \frac{\zeta(\path)}{\mu_{q'}} \right]}{\beta \pr(\hat{A}_j| \emis, \att) } \right\rangle_{(\hat{\A}| J, \emis, \att)} \right\rangle_{( J| \emis, \att)},
%\end{align}
%To perform the averaging operation, we realize that 
Thus, $\pr(\bAhat_j| \emis, \att)$ in the denominator cancels out with the corresponding term in expression for $\pr(\hat{\A}| J, \emis, \att)$ in the numerator. 
Marginalizing over the rest of the variables reduces the second term in Eq.~\eqref{double_diff_loglikelihood_mu} to
\begin{align}
& \left\langle \sum_{j=1}^J \frac{\sum_{\P} \lambda(\P) s_{\eff} (\P)  \left[ -\Delta_q(\P) + \frac{\zeta_q(\path)}{\mu_q} \right] \left[ -\Delta_{q'}(\P) + \frac{\zeta_{q'}(\path)}{\mu_{q'}} \right]}{\beta }  \right\rangle_{( T| \emis, \att)} = \nonumber \\
&  \frac{J}{\beta }\sum_{\P} \lambda(\P) s_{\eff} (\P)  \left[ -\Delta_q(\P) + \frac{\zeta_q(\path)}{\mu_q} \right] \left[ -\Delta_{q'}(\P) + \frac{\zeta_{q'}(\path)}{\mu_{q'}} \right].
% T {\sum_{\P} \lambda(\P) s_{\eff} (\P)  \left[ -\Delta_q(\P) + \frac{\zeta(\path)}{\mu_q} \right] \left[ -\Delta_{q'}(\P) + \frac{\zeta(\path)}{\mu_{q'}} \right]},
\end{align}

The third term in Eq.~\eqref{double_diff_loglikelihood_mu} is independent of event attributes. The measurement time, T is Erlang-distributed random variable with shape $J$ , rate $\beta$ and mean $\frac{J}{\beta}$. Thus the second term is the negative of the third term in Eq.~\eqref{double_diff_loglikelihood_mu} and cancels out, leading to Eq.~\eqref{fim_attn}. 

Performing a similar analysis as above, from Eq.~\eqref{double_diff_loglikelihood_cross} and Eq.~\eqref{double_diff_likelihood_lambda} leads to Eq.~\eqref{fim_cross} and Eq.~\eqref{fim_act} respectively.
%\end{figure*}

%Assuming a uniform distribution of events over all detector angles, this optimization can reduce the number of operations on the scale of angular resolution of paths for voxels, $\{q:\lambda_q\neq0\}$. 

%This step however required knowing the detector bin on which a given path ended, and this term was computed and stored when determining the sensitivity of the collimator. 
%We also used the technique of splitting up the set of all possible paths starting from the $q^{\th}$ voxel into two sets, corresponding to the paths that the unscattered and scattered events followed.
%When processing the paths that the unscattered events follow, we consider only the unscattered events, and likewise, when we analyze the scattered paths, we consider only the scattered LM events. 
%\noindent\begin{minipage}{\textwidth}
\renewcommand\footnoterule{}
\LinesNotNumbered
\begin{algorithm}[h]
%c CODE --------------------------------------------------------------
%\SetStartEndCondition{ (}{)}{)}\SetAlgoBlockMarkers{\{}{\}}%
%\SetKwProg{Fn}{}{}{}\SetKwFunction{FRecurs}{void FnRecursive}%
%\SetKwFor{For}{for}{}{}%
%\SetKwIF{If}{ElseIf}{Else}{if}{}{elif}{else}{}%
%\SetKwFor{While}{while}{}{}%
%\SetKwRepeat{Repeat}{repeat}{until}%
%\AlgoDisplayBlockMarkers\SetAlgoNoLine%
%--------------------------------------------------------------
\DontPrintSemicolon
\SetAlgoLined
\SetAlgoNoLine
\SetKwInput{Input}{Input}
\SetKwInOut{Output}{Output}
\SetKw{Initialize}{Initialize}
\SetKw{PrecalculateStore}{Calculate and Store}
\SetKw{Calculate}{Calculate}
\SetKw{Call}{Call}
\SetKw{Set}{Set}
%% font
\SetKwSty{textbf}
\SetFuncSty{texttt}
%Function Definitions---------------------------
\SetKwFunction{FnSumAjKernel}{K\_SumAjKernel}
\SetKwFunction{FnAjActTermKernel}{K\_AjActTermKernel}
\SetKwFunction{FnAjAttnTermKernel}{K\_AjAttnTermKernel}
\SetKwFunction{FnFisherCalcKernel}{K\_CalcFisherKernel}
\SetKwFunction{CalcRadPath}{CalcRadPath}
\SetKwFunction{CalcDelPath}{CalcDelPath}
%Data Definitions------------------------------
\SetKwData{sysParam}{sysParam}
\SetKwData{lmData}{lmData}
\SetKwData{}{}
\SetKwData{}{}

\Input{$\emis$, $\attn$ ($N$-dimensional arrays), \lmData(LM data containing attributes of $J$ LM events)}%, $\{\A_j:j=1,2,...J\}$}
\KwData{\sysParam (SPECT system geometry)\;
%\quad\quad\quad\ \lmData, which contains List-mode data\;
}
\Output{Fisher Information Matrix, $F_M$ of size $2N \times 2N$}
\BlankLine
\nextnr\Initialize GPU context\;
\nextnr\PrecalculateStore intersection length and corresponding voxel index for different sub-paths in the arrays $\boldsymbol{\Delta}$ and $voxIndex$ respectively\;
\nextnr\PrecalculateStore radiological path, $radPath$ (Eq.~\eqref{eq:pathintegral}) using $\attn,\boldsymbol{\Delta}$ and $voxIndex$\;
\nextnr\Calculate number of voxels with non-zero activity, $N_{nz}$ \;
\nextnr\Set Appropriate grid and block size, and shared memory size for each kernel\;
\BlankLine
\nextnr\tcc{Store $\sum_{\path} \pr(\bAhat_j | \path) \lambda(\path) s_{\eff}(\path)$ for each $j$ in the array $sumAj$}
\Indp$sumAj[j] \leftarrow 0 \text{, for } j=0,1,...,J-1$\;
\For{$i=0$ \KwTo $N_{nz}-1$}{
	\Call{\FnSumAjKernel$<<<$grid1,block1,shMem1$>>>(\emis, \attn, radPath, voxIndex, \boldsymbol{\Delta}, i,$ \sysParam, \lmData, $ sumAj)$}
}
\Indm
\BlankLine
\nextnr\tcc{Calculate $prAjActTerm[q][j]=\left\{ \sum_{P_q} \pr(\bAhat_j | \P) s_{\eff} (\P)  \right\},j=0,1,...,J-1,q=0,1,...,N-1$}
\Indp$prAjActTerm[q][j] \leftarrow 0,j=0,1,...,J-1,q=0,1,...,N-1$\;
\For{$i=0$ \KwTo $N_{nz}-1$}{
	\Call{\FnAjActTermKernel$<<<$grid2,block2,shMem2$>>>(\emis, \attn, radPath, voxIndex, \boldsymbol{\Delta}, i,$ \sysParam, \lmData, $ prAjActTerm)$}
}
\Indm
\BlankLine
\nextnr\tcc{Calculate $prAjAttnTerm[q][j]={ \sum_{P} \pr(\bAhat_j | \P) \lambda(\P)  s_{\eff} (\P)  \left[ -\Delta_q(\path) + \frac{\zeta_q(\path)}{\mu_q} \right] },j=0,1,2,...J-1,q=0,1,2,...N-1$}
\Indp$prAjAttnTerm[q][j] \leftarrow 0,j=0,1,...,J-1,q=0,1,...,N-1$\;
\For{$i=0$ \KwTo $N_{nz}-1$}{
	\Call{\FnAjAttnTermKernel$<<<$grid3,block3,shMem3$>>>(\emis, \attn, radPath, voxIndex, \boldsymbol{\Delta}, i,$ \sysParam, \lmData,$ prAjAttnTerm)$}
}
\BlankLine
\Indm
\nextnr\tcc{Calculate FIM matrix elements one block/chunk at a time}
\Indp$F_M[i][j] \leftarrow 0,i=0,1,2,...,2N-1,j=0,1,2,...,2N-1$\;
\For{$chunk=0$ \KwTo $N_{chunk}-1$}{
\Call{\FnFisherCalcKernel$<<<$grid4,block4,shMem4$>>>(chunk, sumAj, prAjActTerm, prAjAttnTerm, F_M)$}
}
\caption{GPU-accelerated algorithm for calculating FIM terms\label{GPU_FIM}}

\footnotetext{Additional \Call to kernels could be needed for different implementations.}
\end{algorithm}
\section*{Appendix C: Describing the GPU-based implementation to compute the FIM terms}
%\end{minipage}
The developed FIM computation software read the input system configuration and the detected LM data.
For each event, we considered all emission and scattered path as mentioned earlier.
Next, the radiological path for each path and the values of  $\Delta_q(\path)$ for each voxel index were computed and stored.
The quantities $\sum_{\path} \pr(\bAhat_j | \path) \lambda(\path) s_{\eff}(\path)$ were computed and stored for each LM event.
This term appeared in the denominator of Eqs.~\eqref{fim_attn}-\eqref{fim_cross2} and was a function of only the LM event index $j$.

The next step evaluated terms of the form $\sum_{\P_q} \pr(\bAhat_j | \path) s_{\eff}(\path)$ for the $q^{\th}$ voxel, which appeared in the numerator of Eqs.~\eqref{fim_attn}-\eqref{fim_act}. For simplicity, we will denote this terms as $S^{(1)}_{q,j}$
which depends on the paths that start from the $q^{\th}$ voxel and attributes of the $j^{th}$ event.
Thus, this term was calculated only for voxels with non-zero activity and the sum was calculated parallely over only those paths that start from $q^{th}$ voxel and result in detection at angle exactly equal to the detector angle attribute of $j^{th}$ event. The quantity $S^{(1)}_{q,j}$ was then stored as a function of the LM event index
and the voxel index q.

To evaluate the FIM terms with respect to the activity coefficients, in accordance with Eq.~\eqref{fim_act}, for each pair of voxel indices $q$ and $q'$, and for each LM event index $j$, the terms ${ \sum_{\P_q} \pr(\bAhat_j | \path) s_{\eff}(\path)}{}$ and ${\sum_{\P_{q'}} \pr(\bAhat_j | \path) s_{\eff}(\path)}{}$, which have been computed and stored in the previous operations as a function of the LM event index and the voxel index, were multiplied.
The result was divided by the square of $\sum_{\path}\pr(\bAhat_j| \P) \lambda(\P) s_{\eff}(\P)$.
Finally, these terms were summed over the LM events, and this led to FIM terms with respect to the activity distribution. All these operations are done in parallel for a subset of all combination of $q$ and $q'$ where this subset can be defined as a block of elements in the original fisher information.
The FIM terms with respect to the attenuation coefficients and the FIM cross terms were also computed following a very similar procedure, but in accordance with Eqs.~\eqref{fim_attn},~\eqref{fim_cross}, and \eqref{fim_cross2}, respectively. The pseudo-code of the implementation is given in Algorithm \ref{GPU_FIM}.
\section*{References}
\bibliographystyle{IEEEtran}
\bibliography{refs_spect_3}

\end{document}